\documentclass[12pt,preprint]{aastex}
\begin{document}
\title{Constraining the LRG Halo Occupation Distribution using Counts-in-Cylinders}
\author{Beth A. Reid}
\affil{Department of Physics}
\affil{Princeton University, Princeton, NJ 08544}
\email{breid@princeton.edu}
\and
\author{David N. Spergel}
\affil{Department of Astrophysical Sciences}
\affil{Princeton University, Princeton, NJ 08544}
\email{dns@astro.princeton.edu}
\begin{abstract}
The low number density of the Sloan Digital Sky Survey (SDSS) Luminous Red Galaxies (LRGs) suggests that LRGs occupying the same dark matter halo can be separated from pairs occupying distinct dark matter halos with high fidelity.  We present a new technique, Counts-in-Cylinders (CiC), to constrain the parameters of the satellite contribution to the LRG Halo-Occupation Distribution (HOD).  For a fiber collision-corrected SDSS spectroscopic LRG subsample at $0.16 < z < 0.36$, we find the CiC multiplicity function is fit by a halo model where the average number of satellites in a halo of mass $M$ is $\left<N_{sat}(M)\right> = ((M - M_{cut})/M_1)^{\alpha}$  with $M_{cut} = 5.0^{+1.5}_{-1.3} (^{+2.9}_{-2.6}) \times 10^{13} M_{\sun}$, $M_1 = 4.95^{+0.37}_{-0.26} (^{+0.79}_{-0.53}) \times 10^{14} M_{\sun}$, and $\alpha = 1.035^{+0.10}_{-0.17} (^{+0.24}_{-0.31})$ at the 68\% and 95\% confidence levels using a WMAP3 cosmology and $z=0.2$ halo catalog.  

Our method tightly constrains the fraction of LRGs that are satellite galaxies, $6.36^{+0.38}_{-0.39}$\%, and the combination $M_{cut}/10^{14} M_{\sun} + \alpha = 1.53^{+0.08}_{-0.09}$ at the 95\% confidence level.  
We also find that mocks based on a halo catalog produced by a spherical overdensity (SO) finder reproduce both the measured CiC multiplicity function and the projected correlation function, while mocks based on a Friends-of-Friends (FoF) halo catalog has a deficit of close pairs at $\sim 1$ Mpc/$h$ separations.  Because the CiC method relies on higher order statistics of close pairs, it is robust to the choice of halo finder.  In a companion paper we will apply this technique to optimize Finger-of-God (FOG) compression to eliminate the 1-halo contribution to the LRG power spectrum.
\end{abstract}
\section{\bf Introduction}
The Sloan Digital Sky Survey (SDSS; \citet{york/etal:2000}) has recorded the largest sample of Luminous Red Galaxies (LRGs), probing a volume of $\sim 1 \; (h^{-1}$ Gpc$)^3$ out to $z \sim 0.5$ \citep{eisenstein/etal:2001} and making it ideal for studying large scale structure.  Understanding the small-scale relationship between the galaxy and dark matter density fields is essential to extracting the linear matter power spectrum from the galaxy power spectrum, even on very large scales \citep{schulz/white:2006, sanchez/cole:2008}.\\

The Halo Occupation Distribution (HOD) is a popular and useful description of this relationship \citep{seljak:2000,peacock/smith:2000,cooray/sheth:2002}, and can be used to constrain the rate of merging, disruption or evolution in well-defined galaxy populations.  \citet{conroy/ho/white:2007}, \citet{white/etal:2007}, and \citet{wake/etal:2008} have used this framework to constrain the rate at which LRGs merge or are disrupted in clusters.  \citet{brown/etal:2008} combine HOD constraints with luminosity function measurements of red galaxies to deduce that stellar mass build-up in clusters occurs primarily in the satellite galaxies or intracluster light between $z=1$ and $z=0$, while the central galaxies grow only modestly, with $L_{cen} \sim M^{1/3}$.  \citet{zheng/coil/zehavi:2007} constrain stellar mass growth between DEEP2 ($z\sim1$) and SDSS ($z\sim0$) galaxies using the HOD description, and \citet{conroy/etal:2007} employ HOD modeling to illuminate the fate of $z \sim 2$ star-forming galaxies.  Others researchers, e.g., \citet{chen:2007} and \citet{ho/etal:2007}, use the HOD description to study the spatial distribution of satellite galaxies.  \citet{wake/etal:2008} also argue that the small-scale clustering at different redshifts constrains the scatter in halo merger histories which can be compared with predictions of hierarchical models.\\

Several groups have used two and three point statistics \citep{blake/collister/lahav:2007, kulkarni/etal:2007, white/etal:2007, zheng/etal:2008, wake/etal:2008, padmanabhan/etal:2008} as well as galaxy-galaxy lensing \citep{mandelbaum/etal:2006} to constrain the HOD of LRGs.  \citet{ho/etal:2007} have taken a more direct approach and used X-ray determined cluster masses to measure $N_{LRG}(M)$.  Though these analyses were performed on samples with different luminosity and redshift ranges, they offer seemingly conflicting results on both the slope $\alpha$ at the high mass limit of the satellite term $N_{sat} \sim M^{\alpha}$ and the fraction of LRGs that are satellite galaxies.  \citet{ho/etal:2007} find $\alpha \approx 0.6$ when fitting the {\em total} LRG number $N_{tot}(M) \propto M^{\alpha}$, \citet{kulkarni/etal:2007} find $\alpha = 1.4$, \citet{blake/collister/lahav:2007} find $\alpha \sim 2.1 - 2.6$, and \citet{zheng/etal:2008} find $\sim 1.8$ for $\sigma_8 = 0.8$; \citet{kulkarni/etal:2007} report a satellite fraction of $\sim 17\%$, while \citet{blake/collister/lahav:2007}'s redshift slices span 3-8\%, and \citet{zheng/etal:2008} find $5-6\%$ for the LRG subsample studied in this paper.  The most luminous elliptical galaxies in \citet{mandelbaum/etal:2006} have a satellite fraction of $\lesssim 10\%$.\\

The low number density of the SDSS Luminous Red Galaxy (LRG) sample {\em suggests} that LRG pairs occupying the same dark matter halo can be separated from pairs occupying distinct dark matter halos with high fidelity.  In this paper we explore that intuition, and show that one-halo pairs can be identified with $\sim 75\%$ completeness and $\lesssim 27\%$ contamination by simple cuts in the transverse separation $\Delta r_{\perp}$ and LOS separation $\Delta r_{\parallel}$.  Furthermore, these pairs can be grouped together using a Friends-of-Friends (FoF) algorithm to estimate the LRG group multiplicity function.  We apply this technique to a sample of LRGs from SDSS to constrain their HOD.  We find that both the high values of $\alpha \sim 2$ and high satellite fractions reported in previous papers are inconsistent with the $0.16 < z < 0.36$ SDSS LRG group multiplicity function measured here.  In contrast to previous methods which rely on 2 and 3 point statistics to constrain the HOD (as in \citet{kulkarni/etal:2007}), our method probes the HOD more directly by estimating the group multiplicity function from the higher order statistics in the LRG density field in the one-halo dominant regime.\\

We present an overview of the CiC method in \S~\ref{overview} and apply it to an approximately volume limited subsample of SDSS LRGs in \S~\ref{data}, addressing the complications of fiber collisions, incompleteness, and complex angular masks. 
The CiC technique developed here requires calibration on mock galaxy catalogs.  We summarize our $N$-body simulation parameters in \S~\ref{sims}.  \S~\ref{hodmodel} presents the HOD model we employ throughout this analysis and details how we populate our simulations with galaxies.  \S~\ref{CiCtechnique} describes the CiC technique to measure the LRG group multiplicity function and its calibration with simulations.  The HOD parameters are fit using a maximum likelihood analysis explicated in \S~\ref{maximumL}.  
In \S~\ref{results} we present the CiC multiplicity function of our SDSS LRG subsample and describe the relation between the CiC and true group multiplicity functions.
We present the constraints on the HOD parameters and their implications for the fraction of LRGs that are satellites, as well as the mass distribution of halos hosting LRG groups with $n_{sat}$ satellites.  Mock catalogs produced using the CiC maximum likelihood HOD and a spherical overdensity (SO) halo catalog agree with the \citet{masjedi/etal:2006} measurement of $w_p(r_p)$ for this sample when the large scale bias is adjusted with a single parameter for the central galaxy HOD.  In \S~\ref{fofkulksec} we compare these results with a mock LRG catalog based on a FoF halo catalog and with other HOD measurements in the literature.  We show that while the FoF and SO catalogs can both match the observed CiC multiplicity function, the FoF catalog produces mock catalogs with a deficit of halos at 1 Mpc/$h$ that is evident in the projected correlation function.  In \S~\ref{assessCiC} we comment on the strengths and weaknesses of the CiC method and summarize our conclusions in \S~\ref{conc}.\\

Throughout this paper we adopt the \citet{spergel/etal:2007} cosmological parameters used in our simulations to convert redshifts to distances: ($\Omega_m, \Omega_b, \Omega_{\Lambda}, n_s, \sigma_8, h$) = (0.26, 0.044, 0.74, 0.95, 0.77, 0.72).  All distances and separations are in comoving coordinates.
\section{\bf Methodology}
\subsection{Overview of the Method}
\label{overview}
The goal of this section is to measure the group multiplicity function of a subsample of the SDSS LRGs.  For a complete spectroscopic sample covering the full sky (or in a periodic simulation box), our method is as follows:
\begin{itemize}
\item Identify `one-halo' pairs of galaxies.  In what follows, our criteria for two galaxies to be a one-halo pair is $\Delta r_{\perp} \leq 0.8$ Mpc/$h$ and $\Delta r_{\parallel} = 20$ Mpc/$h$ (equivalently $\Delta z/(1+z) \leq \beta_{max} = 0.006$), where both are comoving separations.  These choices were motivated by results on mock LRG catalogs and will be discussed in more detail in later sections.
\item Group pairs of galaxies into groups using a FoF algorithm.  The number of groups with $n_{sat}$ satellites, $N_{CiC}(n_{sat})$ is the CiC multiplicity function.
\end{itemize}
In \S~\ref{data} we present the technical details of accounting for the facts that the SDSS has boundaries and holes, that the spectroscopic sample of LRGs is incomplete, and that the SDSS cannot simultaneously take spectra of two objects separated by $< 55''$, so that regions of the sky observed only once spectroscopically may have missing close pairs of LRGs.  We use the LRGs from the SDSS imaging sample to supplement the spectroscopic sample \citep{blanton/etal:2005, adelman-mccarthy/etal:2007}.  We identify potential pairs from the imaging sample and calibrate this step using pairs of objects from the spectroscopic sample.  Since most nonisolated LRGs are in groups of 2 and candidate pairs from the imaging sample neighboring more than one LRG are highly likely to be group members, we apply the small correction for false LRG pair detections to $N_{CiC}(n_{sat} = 1)$.  A more complex scheme involving corrections at each $n_{sat}$ would not have enough statistics to calibrate on the spectroscopic sample.  We eliminate from our sample LRGs close to the survey boundary, though they are allowed to be grouped with LRGs away from the boundary.  This ensures that our multiplicity function is not biased due to unobserved LRGs outside the boundary.  However, since the bright star masks are numerous and individually very small, this approach is not practical for dealing with objects near bright star masks.  Instead, we adjust $N(n_{sat})$ by estimating the probability that there is an LRG covered by each bright star mask, and then computing the change in $N(n_{sat})$ if there were one.  Table~\ref{table:sdssmulttable} shows that all of these corrections are small.\\

\subsection{\bf Data}
\label{data}
The SDSS \citep{stoughton/etal:2002,abazajian/etal:2003} has imaged $\sim 10^4$ deg$^2$ in $u$, $g$, $r$, $i$, and $z$.  From this sample, spectroscopic LRG targets are efficiently selected using two color/magnitude cuts \citep{eisenstein/etal:2001}.  The tiling algorithm ensures nearly complete samples \citep{blanton/etal:2003}.  However, spectroscopic fiber collisions prohibit simultaneous spectroscopy for objects separated by $<55''$, leaving $\sim 7\%$ of targeted objects without redshifts \citep{masjedi/etal:2006}.  Overlapping plates on $\sim 1/3$ of the survey area mitigate this problem and permit us to calibrate this effect in our analysis, as detailed below.  The `photometric sample' as referred to below consists of objects from the imaging sample that were targeted as LRGs according to the color/magnitude cuts laid out in \citet{eisenstein/etal:2001} but lack spectra.  The `spectroscopic sample' consists of objects from the imaging sample that were targeted as LRGs and subsequently observed.\\

The goal of this analysis is to measure the group multiplicity function for the spectroscopic LRG sample with $-23.2 < M_g < -21.2$ and $0.16 < z < 0.36$.  This sample is approximately volume-limited with $\bar{n} \approx 9.7 \times 10^{-5} (h^{-1}$ Mpc)$^{-3}$, and \citet{zehavi/etal:2005a} and \citet{masjedi/etal:2006} have measured the projected correlation function for this sample on small and intermediate scales.  We begin with the entire set of LRG target galaxies for the DR4$+$ sample from the NYU Value-Added Galaxy Catalog (VAGC) \citep{blanton/etal:2005, adelman-mccarthy/etal:2007} so that our sample {\em includes} the complete sample of galaxies satisfying our $M_g$ and $z$ cuts.  We correct our multiplicity function statistically for the inclusion of a small number of interlopers.\\

Our analysis requires the identification of all close pairs of LRGs in the sample, where close pairs satisfy $\Delta \theta \leq \theta_{max}(z)$ and $\Delta r_{\parallel} \leq \Delta r_{\parallel,max}$.  The dominant source of LRG close pair incompleteness is from LRG-LRG fiber collisions; this results in a deficit of pairs separated by $<55''$ in the spectroscopic sample.  LRG-main galaxy fiber collisions or spectroscopic incompleteness result in further pair incompleteness.  Finally, the bright star mask occupies  $1.88\%$ of the survey area.  We statistically correct for the unobserved LRGs in these regions.\\

Following \citet{eisenstein/etal:2005} we cut sectors with $<60\%$ spectroscopic completeness.  These sectors lie primarily along the boundary of the survey, and remove only 276 objects from our sample.  The remaining survey area is 5564 deg$^2$, of which 104 deg$^2$ is covered by a bright star mask.  Using inverse random catalogs from the VAGC to trace the survey geometry, we remove all objects within $\theta_{max}(z_{min} = 0.16)$ of the survey boundary, which amounts to 469 deg$^2$, or 8.6\% of the total area.  However, we keep track of all LRGs in the boundary, because they are allowed to form pairs with the objects remaining in our sample and therefore contribute to the group multiplicity function.  Only 53 objects from the boundary were included in CiC groups.  Note that while there are many corrections for incompleteness, they are all small and well-constrained.\\

\subsubsection{Sample Model and Color/Magnitude cuts}
\label{photosamplemodeling}
\begin{figure}
\includegraphics*[scale=0.75,angle=0]{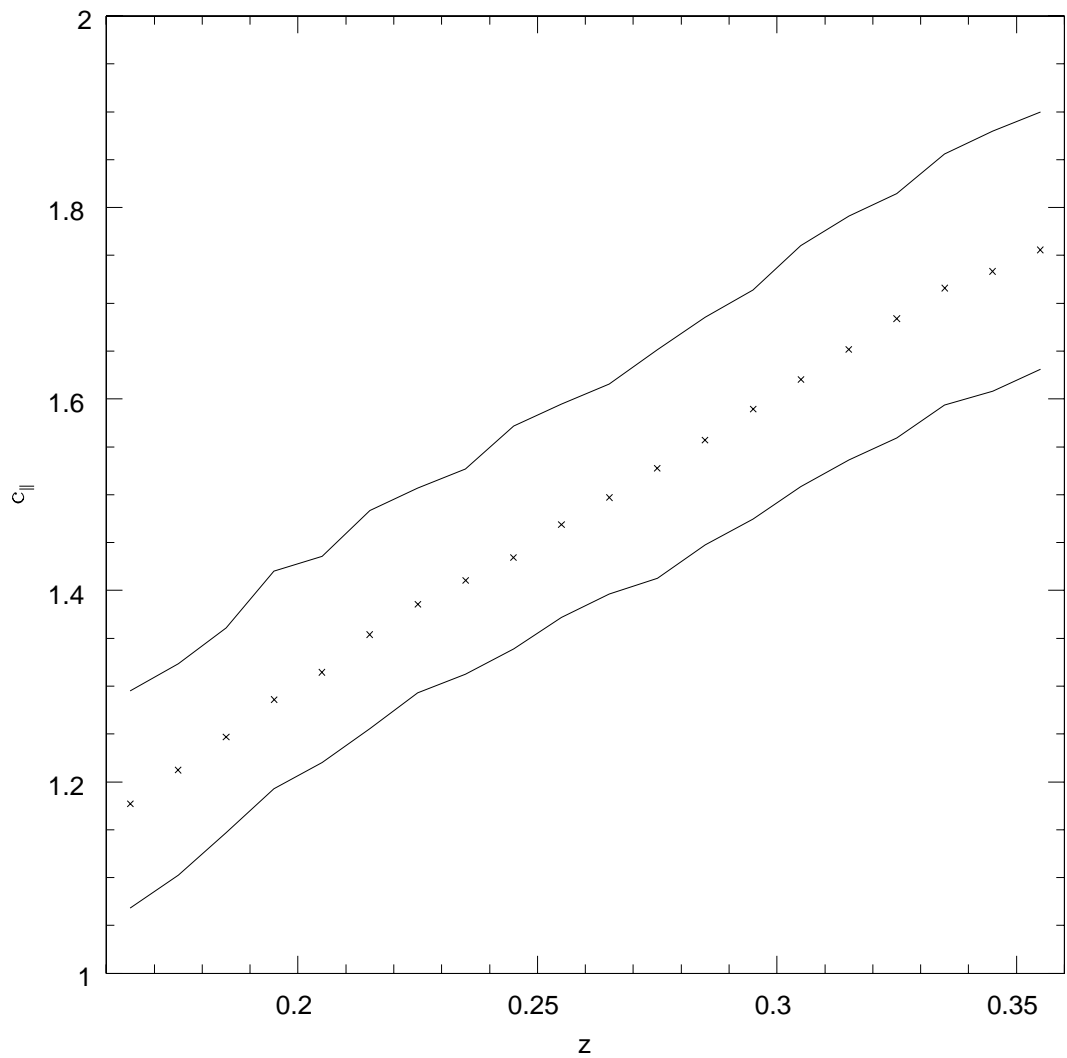}
\caption{\label{fig:cparzbands} The mean $c_{\parallel}-z$ relation and bands including 95\% of objects in each $\Delta z = 0.01$ bin.  $c_{\parallel}$ (Eqn.~\ref{cparcuteqn}) is used as a redshift indicator for objects targeted as LRGs but lacking spectra.} 
\end{figure}

There are 41721 objects in our spectroscopic sample passing the redshift, $M_g$, sector completeness, and boundary cuts, and 8167 objects from the photometric sample passing the same sector completeness and boundary cuts.  To reduce the level of contamination in our LRG multiplicity function, we apply color and magnitude cuts to objects from the photometric sample falling within $\Delta \theta_{max}$ of a spectroscopic LRG to select a high-fidelity sample of close pair candidates.  Since we know the redshift of the proposed group from the spectroscopic neighbor, we use $c_{\parallel}$ \citep{eisenstein/etal:2001} as a redshift indicator and $r_{pet}$ of the photometric sample object as an absolute magnitude indicator at the spectroscopic redshift.
\begin{equation}
\label{cparcuteqn}
c_{\parallel} = 0.7(g - r) + 1.2(r - i - 0.18)
\end{equation}
Using the colors of objects in our spectroscopic sample, we find the upper and lower limits of $c_{\parallel}$ that encompass 95\% of the spectroscopic LRGs as function of redshift in bins of $\Delta z = 0.01$.  This relation is shown in Fig.~\ref{fig:cparzbands}.  For a candidate close pair match, we discard photometric sample objects with $c_{\parallel}$ falling outside this region, given the redshift of the spectroscopic group member.
Since the k+e corrections described in \citet{eisenstein/etal:2001} have already been applied to the spectroscopic sample, we estimate the expected $M_g$ of a photometric object at redshift $z$ of the spectroscopic group member by
\begin{equation}
\label{magcuteqn}
M_{g,photo} = M_{g,spec} + (r_{pet,photo} - r_{pet,spec})
\end{equation}
and disregard objects with $M_{g,photo} < -23.2$ or $M_{g,photo} > -21.2$.\\

For a random sample of LRG targets, 36.6\% of them will pass the \citet{eisenstein/etal:2001} LRG cuts as well as our sample's magnitude and redshift cuts, based on the ratio in DR4$+$ for targeted LRGs that have spectra.  The fraction of objects in our photometric sample that would pass our spectroscopic magnitude and redshift cuts will be slightly higher than for the random sample of LRG targets, since it includes fiber collision pairs very likely to be LRGs.  We compute this fraction for the photometric sample at the end of the group-finding algorithm to circumvent this issue.  We model the photometric sample as composed of $N_{pass}$ objects which would pass the absolute magnitude and redshift cuts of our sample, if they had spectra; the remaining objects we label as $N_{fail}$.
A large fraction of objects in the $N_{fail}$ group are below our minimum redshift cut $z_{min} = 0.16$.  Tests on targeted LRGs with spectra that fail our LRG subsample cuts show that the $N_{fail}$ group is well-approximated as uncorrelated with the spectroscopic LRG sample.\\

Candidate close pairs of spectroscopic LRGs from the photometric sample are naturally divided into two groups.  The first, denoted `FB' for `fiber', includes objects $\theta \leq 55''$ from a spectroscopic object.  These objects do not have spectra primarily due to a fiber collision with the neighboring spectroscopic LRG.  The bulk of candidate close pairs fall into this category, and the contamination for such pairs by objects at different redshifts is low.  The remaining close pairs have separations $55'' < \theta < \theta_{max}(z_{spec})$, arise from the small overall incompleteness of the survey or fiber collisions with MAIN galaxies, and are denoted `INC'.  The probability of contamination is larger for this type, but still manageable.  We use all targeted LRGs with spectra but failing our $M_g$ or z cuts to compute an average rate at which an $N_{fail}$ object will pass our color/magnitude cut on the photometric sample by comparison with the spectroscopic LRG pairs in each type of collision: $p_{pass,FB} = 0.078$ and $p_{pass,INC} = 0.070$.  The rate is slightly lower for INC collisions because they are more likely to be at a lower redshift where $c_{\parallel}$ is more discriminating, since $\theta_{max}(z)$ decreases with $z$.  The number of close pair contaminants we expect from this sample for our two types of collisions are
\begin{eqnarray}
N_{FB,fail} = \frac{A_{FB}}{A_{survey}}*N_{fail}*p_{pass} \approx 2.1 \label{failcontamFB}\\
N_{INC,fail} = \frac{A_{LRG} - A_{FB}}{A_{survey}}*N_{fail}*p_{pass} \approx 26.8 \label{failcontamINC}
\end{eqnarray}
where $A_{FB}$ and $A_{LRG}$ are the total areas enclosed by annuli of 55'' and $\theta_{max}(z_{spec})$ around each LRG, and $A_{survey}$ is the total survey area after removal of the boundary.  $N_{fail}$ was estimated at the end of the group-finding algorithm; see below.\\

To estimate the completeness and contamination for the $N_{pass}$ sample, we rely on observed collisions from the spectroscopic sample.  For FB collisions, we found 295.5 collisions between objects in the spectroscopic sample that passed the color/magnitude cuts.  276 of these passed the $\Delta z$ cut, so 93.4\% of objects satisfying $\Delta \theta \leq 55''$ and passing the color/magnitude cuts would be considered one-halo pairs if the redshift were known.  Furthermore, 19 pairs failed the color/magnitude cuts but passed the $\Delta z$ cut.  Therefore, the sample is 93.5\% complete after the color/magnitude cuts and 6.6\% contaminated.  For INC collisions, we found 2494 collisions passing the photometric color/magnitude cuts; 2020 of these passed the $\Delta z$ cuts, so would be assigned as pairs.  110.5 pairs passing the $\Delta z$ cuts failed the color magnitude cuts.  Therefore, this collision sample should be 94.8\% complete and 19.0\% contaminated.\\

Our approach is to estimate the number of interlopers and apply the correction to the number of groups of 2 LRGs.  568 objects from the photometric sample are $\leq 55''$ from a spectroscopic LRG object and pass the color/magnitude cuts at the spectroscopic redshift; 250 objects from the photometric sample are $55'' < \theta < \theta_{max}(z)$ and pass the color/magnitude cuts at the spectroscopic redshift.  Based on the rates measured from spectroscopic collisions and after correcting for the expected number of interlopers $N_{FB,fail} + N_{INC,fail}$, we expect 709.3 of the remaining 789.1 collisions to be `true' collisions (i.e., would pass the $\Delta z$ criterion as well, if the redshift were measured), and we expect to have missed 46.3 true collisions due to our color/magnitude cuts.  Therefore, we overestimate the total number of pairs by 62.4 (a 2.8\% correction).\\

Finally, we must estimate the total number of isolated LRGs (i.e., those with no neighboring LRGs passing our CiC cuts) missing from our sample due to the incompleteness of the spectroscopic sample.  Again making use of the assumption that objects in the photometric sample that would fail the color/magnitude cuts of our LRG sample are uncorrelated with the spectroscopic sample and that the number of expected interlopers is negligible (Eqns.~\ref{failcontamFB} and ~\ref{failcontamINC}), we expect the total number of isolated galaxies in the photometric sample to be
\begin{equation}
\label{isolatedcorrection}
N_{photo,iso} = N_{pass} (1-p_{group}) + N_{fail},
\end{equation}
where $p_{group}$ is the probability that an LRG is in a CiC group with $n_{sat} \geq 1$.  Using the observable $N_{photo,iso}$, the number of objects from the photometric catalog not grouped with spectroscopic objects, along with the pass rate for a random set of LRG targets of 36.6\%, allows us to solve for the ratio $N_{pass}/N_{fail}$ in the photometric sample.  Given the final estimate of the group multiplicity function (see below), we find $p_{group} = 0.132$ and the isolated LRG contribution from the photometric sample, $N_{pass}*(1-p_{group})$, is 2900.  This produces an increase in total objects in our sample of 8\%, in agreement with the incompleteness rate reported by \citet{masjedi/etal:2006}.\\

\subsubsection{Group-finding Algorithm and Bright Star Mask Corrections}
\label{datagroupfinding}
One-halo pairs of galaxies are assigned by satisfying the criteria $\Delta r_{\perp} \leq r_{\perp,max}$ and $\Delta z/(1+z) \leq \beta_{max}$.  Galaxies are then grouped by a FoF algorithm.  The challenges in this section are to self-consistently incorporate the photometric sample into the calculation of the FoF group multiplicity function, and to marginalize over tiny Bright Star Mask holes in the survey geometry.\\

The group finding algorithm first assigns candidate pairs for each galaxy.  As described in \S~\ref{photosamplemodeling}, the redshift space criterion for a spectroscopic-photometric pair becomes $c_{\parallel,min}(z_{spec}) \leq c_{\parallel,photo} \leq c_{\parallel,max}(z_{spec})$ and $-23.2 < M_{g,photo} < -21.2$.  Using these pair assignments, candidate groups are formed with the FoF algorithm.  Spectroscopic neighbors of photometric objects in the group are kept if they satisfy the $\Delta r_{\parallel}$ requirement with at least one spectroscopic object in the group, and photometric neighbors of photometric objects in the group are kept if they satisfy the redshift space criterion with at least one spectroscopic object in the group.  We correct the group assignments for the 23 photometric objects in two distinct groups.  We first give preference to groups in which the photometric object is a fiber-collision pair; otherwise, we assign the photometric object to the larger group.  We must then recompute the membership of the group losing the photometric object.\\

The last step of the group-finding algorithm accounts for Bright Star Mask and counts the number of isolated photometric objects, $N_{photo,iso}$ used in Eqn.~\ref{isolatedcorrection}.  The Bright Star Mask catalog, consisting of the {\bf ra}, {\bf dec}, and radius for each circular masked region, was generated from the Tycho2 catalog as described by the NYU VAGC \citep{blanton/etal:2005}.  The Bright Star Mask covers 1.88\% of the total survey region.  Using the observed pair counts as a function of radius, we estimate the probability per unit area of finding a 1-halo pair satisfying $r_{\perp} \leq r_{\perp,max} = 0.8$ Mpc/$h$ and $\Delta z/(1+z) \leq \beta_{max}$ as a function of $r_{\perp}$.  The result is well-fit with a power law: $dP/dA = 0.027 r_{\perp}^{-1.08}$ (Mpc/$h)^{-2}$.  $dP/dA$ is similar to the projected correlation function, except that it does not subtract the uncorrelated contribution, and it only includes pairs meeting our $\Delta r_{\parallel}$ criterion.  For each bright star mask within $r_{\perp,max}$ of an LRG, we evaluate the probability that an unobserved LRG resides under the Bright Star Mask by integrating our power law fit to $dP/dA$ over the area of overlap between the Bright Star Mask and the LRG annulus satisfying $r_{\perp} \leq r_{\perp,max}$.  We assign the probability of a given Bright Star Mask to cover an LRG as the maximum probability assigned from all the observed LRG neighbors of the Bright Star Mask elements.  We modify the final group multiplicity function to account for each possible unobserved LRG, including its ability to bridge two previously existing groups.  Therefore a single group of LRGs can contribute to many group multiplicity bins with weights according to the probability of a covered LRG beneath each Bright Star Mask intersecting a group member.  This procedure adds 79.8 net nonisolated LRGs to the group multiplicity function; the rest of the expected $0.0188*N_{\rm LRG,observed}$ are added to the isolated LRG count.  See Table~\ref{table:sdssmulttable} for the final multiplicity function along with the contributions of each of these corrections.\\ 

Finally, we place an upper limit on the contribution from photometric-object only groups.  As listed in the final column in Table~\ref{table:sdssmulttable}, we find 64 photometric object pairs and a single group of 3 photometric objects for which the $c_{\parallel}$ and $r_{petro}$ for each galaxy in the group fall into the 95\% range of the same redshift bin of Fig.~\ref{fig:cparzbands}.  30\% of such pairs fall into the $z = 0.35 - 0.36$ bin, for which $c_{\parallel}$ is the least discriminating, while only 7.8\% of the spectroscopic LRGs lie in this redshift range.  Therefore, we conclude that we miss a negligible number of groups by ignoring photometric object only groups.\\
\begin{deluxetable}{lllllll}
\tabletypesize{\scriptsize}
\tablewidth{0pt}
\tablecolumns{7}
\tablecaption{\label{table:sdssmulttable} This table lists the CiC group multiplicity function of the SDSS LRG spectroscopic $+$ photometric sample for $\Delta r_{\perp,max} = 0.8 Mpc/$h and $\beta_{max} = 0.006$.  Column 1 lists the number of satellites in the group, $n_{sat} = n_{group} - 1$.  Column 2 contains the raw group multiplicity before any corrections have been made (including for double counting of photometric objects).  Column 3 lists the final group multiplicity function after all the corrections of \S~\ref{datagroupfinding}.  Column 4 shows the number of photometric objects in groups with $n_{sat}$ satellite galaxies, Column 5 shows the net effect of the Bright Star Mask correction, and Column 6 shows the number of groups of photometric objects only with $n_{sat}$ satellites satisfying the criteria listed in \S~\ref{datagroupfinding}.  These groups are largely false detections and do not contribute to the counts in Column 2.  The total area of the survey, excluding the boundary region but including the bright star mask area, is 1.549 steradians.}
\tablehead{
\colhead{$n_{sat}$} & \colhead{$N_{\rm raw}(n)$} & \colhead{$N_{\rm final}(n)$} & \colhead{$n_{\rm photo}$} & \colhead{$\Delta N_{\rm BSMASK}$} & \colhead{$N_{\rm photo, only}$} & \colhead{$N_{CiC,mocks}(n)$}}
\startdata
0 & 36778 & 40407.56 & 2907.71 & 702.49 & 0 & 40546.1\\
1 & 2314 & 2301.12 & 612 & 63.54 & 64 & 2190.4\\
2 & 281 & 285.86 & 146 & 6.86 & 1 & 323.2\\
3 & 56 & 54.29 & 38 & 1.29 & 0 & 65.8\\
4 & 20 & 20.20 & 14 & 0.20 & 0 & 15.2\\
5 & 4 & 4.13 & 5 & 0.13 & 0 & 4.7\\
6 & 1 & 1.05 & 2 & 0.05 & 0 & 1.5\\
7 & 1 & 1.02 & 1 & 0.02 & 0 & 0.2\\
8 & 0 & 0.02 & 0 & 0.02 & 0 & 0.16\\
\enddata
\end{deluxetable}
\subsubsection{Note on average $\bar{n}_{LRG}$}
\citet{zehavi/etal:2005a} construct a model of the spectroscopic LRG sample and estimate $\bar{n}_{LRG} = 9.7\times10^{-5}$ (Mpc/$h)^{-3}$ for the $0.16 < z < 0.36$, $ -23.2 < M_g < -21.2$ subsample.  This is in good agreement with our estimate $N_{LRG}/V = 1.0\times 10^{-4}$ (Mpc/$h)^{-3}$ where V is the product of the fraction of the sky covered by our sample and the comoving volume between $z=0.16$ and $z=0.36$.  As discussed in the previous sections, the number of LRGs in our sample is increased by 8\% when including objects from the imaging sample and covered by Bright Star Masks.  For our choice of cosmological paramaters, the volume of the $0.16 < z < 0.36$ shell is 5\% larger than in the \citet{zehavi/etal:2005a} cosmology, so a difference of 3\% in $\bar{n}_{LRG}$ is expected.\\
\clearpage

\subsection{\bf Simulations and Halo Catalogs}
\label{sims}
In order to constrain the LRG HOD, we require mock LRG catalogs to calibrate the relationship between the CiC group multiplicity function and the true HOD multiplicity function.  The mock LRG catalogs are derived from halo catalogs from a $z=0.2$ $N$-body simulation snapshot using the HOD formalism detailed in \S~\ref{hodmodel}.  We use a $1024^3$ particle TPM \citep{bode/ostriker:2003} 1 $(h^{-1} \; {\rm Gpc})^3$ simulation with particle mass $M_{p} = 6.72 \times 10^{10} M_{\sun}$ and $\epsilon = 16.28  \; h^{-1} \; {\rm kpc}$  described in more detail in \citet{sehgal/etal:2007}.  The cosmological parameters are set to ($\Omega_m, \Omega_b, \Omega_{\Lambda}, n_s, \sigma_8, h)$ = (0.26, 0.044, 0.74, 0.95, 0.77, 0.72).  For our FoF catalogs \citep{davis/etal:1985}, all halos containing $\geq 107$ particles were identified using a linking parameter $b = 0.2$.  The virial mass and radius for each halo are measured in spheres with overdensity determined by the spherical top hat collapse model \citep{bryan/norman:1998}.  We also find the radius $r_{v_{max}}$ at which the circular velocity is maximum in order to estimate the halo concentration $c_{vir}$.  Eqn. 11 in \citet{lokas/mamon:2001} implies
\begin{equation}
\label{conceqn}
c_{vir} = 2.16 r_{vir}/r_{v_{max}}.
\end{equation}
We produce catalogs using the spherical overdensity (SO) halo finder with $\Delta = 200\rho_b$  described in \citet{tinker/etal:2008} and code kindly provided by J. Tinker.  We record halos down to 50 particles.  \citet{reid:phd} shows that the clustering and velocity statistics of halos containing 50-80 particles are not altered by the mass resolution.  The mass function in this mass range is overestimated by $\sim 5\%$ and unaffected in higher mass bins.  As we demonstrate in later sections, only a small fraction of LRGs occupy halos in the lowest mass bin for our best fit HOD, so the effect of simulation resolution on the HOD parameters should be minimal.\\

\subsection{\bf HOD model}
\label{hodmodel}
The Halo Occupation Distribution (HOD) model assumes that the probability $P(N_{LRG}|M)$ of $N_{LRG}$ LRGs occupying a dark matter halo of mass $M$ at redshift $z$ depends only on the halo mass (for a review, see \citet{cooray/sheth:2002}).  However, the application of the HOD formalism requires that we make several further assumptions about $P(N_{LRG}|M)$.  Detailed studies of dark matter halos and subhalos suggest a division of galaxies into central and satellite galaxies \citep{kravtsov/etal:2004}.
  The central galaxies are assumed to sit at the halo center, consistent with the observation that most ($\sim 80\%$) of the brightest cluster LRGs are found within $0.2r_{vir}$ of the center of the cluster potential well as traced by X-rays.  Satellite galaxies occur in the more massive halos already containing a central galaxy.  In high resolution simulations they can be directly associated with dark matter subhalos \citep{vale/ostriker:2006}; here we will assume they have the same distribution as the halo dark matter.  
We will use these functional forms with the five free parameters $M_{min}$, $\sigma_{log M}$, $M_{1}$, $M_{cut}$, and $\alpha$ to describe the mass dependence of the average halo occupation as a function of halo mass $M$:
\begin{eqnarray}
\left<N(M)\right> = \left<N_{cen}\right>(1 + \left<N_{sat}\right>) \label{censatsum}\\
\left<N_{cen}\right> = \frac{1}{2} \left[ 1 + {\rm erf} \left ( \frac{log_{10} M - log_{10} M_{min}}{\sigma_{log M}}\right)\right]\label{NcenM}\\
\left<N_{sat}\right> = \left(\frac{M - M_{cut}}{M_1}\right)^{\alpha}\label{NsatM}
\end{eqnarray}
This form for $N_{cen}(M)$ was suggested in \citet{zheng/etal:2005} and adopted in the analysis of \citet{blake/collister/lahav:2007}.  Both \citet{blake/collister/lahav:2007} and \citet{kulkarni/etal:2007} use Eqn.~\ref{NsatM} with $M_{cut} = 0$.  As shown below, we find evidence for $M_{cut} > 0$.  Half of halos with mass $M_{min}$ host central LRGs, and $\sigma_{log M}$ quantifies the width of the transition from $N_{cen}(M) = 0$ to $N_{cen}(M) = 1$.  $M_1$ sets the mass scale at which satellite galaxies become probable, and $M_{cut}$ sets a cut-off below which halos do not host satellites.  In the case $M \gg M_{cut}$ and $M_1 \gg M_{min}$, we expect $\alpha = 1$, or the number of satellite galaxies to be proportional to the halo mass.\\

$P(N_{LRG,sat}|N(M))$ is assumed to be Poisson distributed; this assumption is supported by the distribution of subhalo counts in simulations \citep{kravtsov/etal:2004} as well as observations \citep{lin/mohr/stanford:2004,ho/etal:2007}.  Often two-point statistics are used to fit the HOD; they depend only on the first and second moments of $P(N_{LRG}|M)$.  The technique we present here to constrain the HOD, Counts-In-Cylinders, makes use of higher-order statistics in the galaxy distribution.  This technique has excellent constraining power for the parameters of $N_{sat}$, but is dependent upon the accuracy of the Poisson assumption for deriving accurate HOD parameters.  Using the Poisson assumption, the expected number of halos with $N_{sat} = n$ satellites is given by 
\begin{equation}
\label{satexp}
\left< N(N_{sat} = n) \right> = \int dM \;n_{halo}(M) \; \exp(-N_{sat}(M) N_{cen}(M)) \; \frac{\left(N_{sat}(M) N_{cen}(M)\right)^n}{n!}.
\end{equation}
Eqn.~\ref{satexp} is central to our maximum likelihood analysis described in \S~\ref{maximumL}.\\

The CiC method constrains only the HOD parameters in $N_{sat}(M)$.  In this work we fix the $N_{cen}$ parameters $M_{min}$ and $\sigma_{log M}$ by matching the observed $\bar{n}$ and amplitude of the projected correlation function in the 2-halo regime.\\

\subsubsection{Populating the Simulations}
Halos are populated with a central galaxy with probability $N_{cen}(M)$.  Central galaxies are placed at the center of their host halos and assigned the peculiar velocity of their halos.  Halos with a central galaxy are populated with $N_{sat}$ galaxies, where $P(N_{sat}|N(M))$ is drawn from a Poisson distribution.  Our parameter constraints are derived using SO halo catalogs.  For those, the position and velocity of the satellite galaxies are taken to be that of a randomly selected dark matter particle halo member.  For mock catalogs based on the FoF halos, the satellite galaxies are independently distributed following an NFW profile with concentration of the dark matter halo determined by Eqn.~\ref{conceqn}.  The peculiar velocity of a satellite galaxy is the sum of the halo peculiar velocity and a random velocity drawn from a Gaussian distribution determined by the virial velocity of the halo \citep{lokas/mamon:2001}:
\begin{equation}
\sigma^2_{vir,1D} = \frac{GM_{vir}}{2R_{vir}}.
\end{equation}
We assign comoving redshift space position $s$ to an object in our mock catalogs using the conversion at $z_{box} = 0.2$:
\begin{equation}
s = x_{LOS} + (1+z_{box})v_{p}/H(z_{box})
\end{equation}
where $x_{LOS}$ is the comoving distance along the line of sight in real space.\\

\subsection{\bf Counts-In-Cylinders Technique}
\label{CiCtechnique}
The hypothesis underlying the Counts-In-Cylinders technique to constrain the LRG HOD is that 1-halo and 2-halo LRG pairs are separable based on their relative angular and redshift space positions.  In the regime of small separations where the 1-halo term dominates, a cylinder should be a good approximation to the density contours surrounding central galaxies, as long as the satellite velocity is uncorrelated with its distance from the halo center, and the relative velocity dominates the separation of central and satellite objects in the redshift direction.  Based on our initial analysis of completeness and contamination of mock catalogs derived from FoF halos, we set $\Delta r_{\perp,max} = 0.8$ Mpc/$h$ and $\Delta z_{max} = 20$ Mpc/$h$ for our $z=0.2$ catalogs.  $\Delta r_{\perp,max}$ is set by the typical comoving size of halos hosting satellite galaxies, and $\Delta z_{max}$ is set by the amplitude of the velocity dispersion in halos massive enough to host satellite galaxies.  In later work we plan to improve the fidelity of our CiC group identification.  However, the choice made here is sufficient since we calibrate the relatively small bias in the method using mock catalogs.\\

Finally, we must select a redshift dependence for the parameters $\Delta r_{\perp,max}$ and $\Delta r_{\parallel,max}$.  The virial radius of a halo of fixed mass in {\em comoving} coordinates decreases by $< 6\%$ over our SDSS sample redshift range; this decrease will be offset as massive halos hosting LRGs accrete and grow.  We approximate the net result by fixing the transverse separation in comoving coordinates.  The virial velocity of a halo of fixed mass varies by a similar fraction, but again we expect it to grow with halo mass.  We keep the effective maximum relative velocity fixed as we translate the cylinder parameters to different redshifts.  For an object at $\chi_o = \chi(z_d)$, the observed redshift will be 
\begin{equation}
z_{obs} = z_d + (1+z_d)v_p/c
\end{equation}
so for pairs in massive halos where $\Delta z_d$ is relatively small, $\Delta z_{obs}/(1+z_{obs}) \approx \Delta v_{p}$/c.\\
\subsection{\bf Maximum Likelihood Parameter Estimation}
\label{maximumL}
While the main purpose of this technique is to produce mock catalogs with higher-order statistics in agreement with the observed sample (specifically the CiC multiplicity function), in the process we derive constraints on HOD parameters.  These results are dependent on the input mass function, which we take directly from our simulation. 
The HOD model we are constraining predicts the expected number of groups containing $N_{sat}$ satellite LRGs for each positive integer $n = N_{sat}$ via Eqn.~\ref{satexp}.  The CiC technique presented here produces a group multiplicity function which may have both an offset and scatter around the true number of groups with $n_{sat}$ satellites,
\begin{equation}
\label{deltag}
\Delta g(n_{sat}) = N_{CiC}(n_{sat}) - N_{true}(n_{sat}),
\end{equation}
where by `true' groups we mean groups of LRGs occupying the same halo.  The overall offset reflects both the incompleteness of CiC method for finding pairs of LRGs in the same halo, as well as the contamination from LRGs in other halos.  The scatter $\Delta g(n_{sat})$ arises both from the stochasticity of the $N_{LRG}-M_{halo}$ relation and cosmic variance of halo alignments that cause pairs of galaxies in separate halos to be grouped together by CiC.  Using our mock catalogs we have estimated $P(\Delta g(n_{sat}))$ for each $n_{sat}$; see \S~\ref{CiCoffsetResults} for more details.\\

While in principle $\Delta g(n_{sat})$ should be measured at each HOD point, this is not feasible since one must make many mock catalogs in order to estimate the distribution of $\Delta g(n_{sat})$.  As we show below, the tightest parameter constraints from our CiC group multiplicity measurement are on $M_1$ and $\alpha + M_{cut}/(10^{14} M_{\sun})$.  Examination of $\Delta g(n_{sat})$ over a subset of points around the maximum likelihood HOD indicates that the variance of $\Delta g(n_{sat})$ is much larger than any systematic variation of $\Delta g(n_{sat})$ with HOD parameters for $n_{sat} = 2,3$.  There may be slight trends at larger $n_{sat}$.  This may be because the CiC parameters adopted in this paper miss a significant fraction of large groups, so that a better model for the offset between $N_{CiC}$ and $N_{true}$ would be $N_{CiC}(n_{sat}) = N_{true}(n_{sat})*b(n_{sat})$.  We chose not to adopt this parametrization because at large $n_{sat}$, either multiplicity function can be 0 and lead to ambiguities in $b(n_{sat})$.  We hope to improve the accuracy of the CiC technique at $n_{sat} > 2$ in future work.  However, we expect this effect to have only a small impact on the final parameter constraints.\\

We assume that for each value of $n_{sat}$, the number of groups with $n_{sat}$ satellites, $j = N_{CiC}(n_{sat}) - \Delta g(n_{sat})$, is Poisson-distributed about the expectation value $\mu_{n}$ in Eqn.~\ref{satexp}:
\begin{equation}
\label{poissondist}
P(j = N_{CiC}(n_{sat}) - \Delta g(n_{sat}) | \mu) = e^{-\mu_{n}} \frac{\mu_{n}^{j}}{j!}
\end{equation}
and independent of the other observed and expected multiplicities.  Next we marginalize over the distribution of $\Delta g(n_{sat})$ which we measured from our mock catalogs:
\begin{equation}
\label{margdeltag}
P(N_{CiC}(n_{sat})| \mu_n) = \int P(j = N_{CiC}(n_{sat}) - \Delta g(n_{sat}) | \mu_n) P(\Delta g) d\Delta g.
\end{equation}
Then the probability of observing the set of CiC multiplicities $\vec{N}_{CiC} = \{N_1, N_2, ... N_{k_{max}} \}$ given the expectation values $\vec{\mu} = \{\mu_1, \mu_2, ... \mu_{k_{max}} \}$  (with the subscript denoting the number of satellites in the group) is
\begin{equation}
\label{independentP}
P(\vec{N}_{CiC} | \vec{\mu}) = \Pi_{n=1}^{k_{max}} P(N_n | \mu_n)
\end{equation}
In order to make use of Bayes' Theorem to find the parameters $\vec{p}$ that maximize the likelihood of the observations $\vec{N}_{CiC}$,
\begin{equation}
\label{MLeqn}
P(\vec{p} | \vec{N}_{CiC} ) = P(\vec{N}_{CiC} | \vec{p}) P(\vec{p})/P(\vec{N}_{CiC})
\end{equation}
we must assume a prior $P(\vec{p})$.  We choose a flat prior on $\alpha$, $M_1$, and $M_{cut}$, except that they are required to be positive.\\
Finally, we note that the $N_{CiC,SDSS}(n_{sat}) = 0$ for $n_{sat} > 9$.  In Eqn.~\ref{independentP} we set $k_{max}=18$.  The exact cutoff is unimportant since $n_{sat} \leq 9$ multiplicities already constrain the HOD parameter space to one in which $N(n_{sat})$ is sharply falling.\\
In Figures~\ref{fig:avgHOD} through ~\ref{fig:dNndlogM} we compute the expectation of several quantities as 
\begin{eqnarray}
\left<y\right> & = \int y(\vec{p}) P(\vec{p} | \vec{N}_{CiC}) d\vec{p} \label{avgHODeqn} \\
\Delta x & = \sqrt{\left<(x - \left<x\right>)^2 \right>} \label{varHODeqn}.
\end{eqnarray}
Eqn.~\ref{varHODeqn} is used to compute the error bars in those figures.  Confidence intervals for some function $y(\vec{p})$ of the HOD parameters given in \S~\ref{HODconstraints} are computed directly from the marginalized distribution:
\begin{equation}
P(y') dy' = \int_{y(\vec{p}) \in [y',y'+dy']}  P(\vec{p} | \vec{N}_{CiC}) d\vec{p}.
\end{equation}
\section{\bf Results}
\label{results}
We use the SDSS LRG $N_{CiC}$ to evaluate the likelihood of a given HOD model using Eqn.~\ref{MLeqn}.  However, $M_{min}$ and $\sigma_{log M}$ are still unconstrained.  We choose $\sigma_{log M} = 0.7$ for the analysis that follows, and set $M_{min}$ so that the number density of LRGs in the mock catalog matches that of our SDSS sample, $\bar{n}_{LRG} = 1.0 \times 10^{-4}$ (Mpc/$h)^{-3}$.  This choice of $\sigma_{log M} = 0.7$ provides excellent agreement with the observed projected correlation function; see \S~\ref{projcorr} and Figure~\ref{fig:wprp1}.\\

\subsection{SDSS LRG $N_{CiC}$}
Column 3 of Table~\ref{table:sdssmulttable} presents our final estimation of $N_{CiC}(n_{sat})$ for $n_{sat} = 0 - 8$, the group multiplicity function for our $0.16 < z < 0.36$, $-23.2 < M_g < -21.2$ subsample of SDSS LRGs.\\ 

\subsection{CiC offset $\Delta g(n_{sat})$ and Group-Finding Accuracy}
\label{CiCoffsetResults}
To characterize the offsets $\Delta g(n_{sat})$ (Eqn.~\ref{deltag}) between the CiC group multiplicity function and the true multiplicity function, we produce 600 mock catalogs using the HOD parameters $\sigma_{log M} = 0.7$, $M_{min} = 8.05 \times 10^{13} M_{\sun}$, $M_{cut} = 4.66 \times 10^{13} M_{\sun}$, $M_1 = 4.95 \times 10^{14} M_{\sun}$, and $\alpha = 1.07$, which are close to the final maximum likelihood parameters reported in \S~\ref{HODconstraints}; the two produce very similar CiC group multiplicity functions.  We measured $\Delta g(n_{sat})$ in a randomly selected cubic subsample of each mock catalog, selected to have a volume equal to that of our SDSS LRG subsample, 0.46 (Gpc/$h)^3$.  The width of $P(\Delta g)$ may be underestimated, since each subsample is drawn from the same dark matter simulation.  However, the broadness of $N_{cen}(M)$ means that our simulation box contains many more halos than LRGs, and so the variance resulting from cosmic structures is somewhat sampled.  The average and variance of $\Delta g(n_{sat})$ are reported in Table~\ref{table:deltagvar}.  Comparison to the observed $N_{CiC}(n_{sat})$ in Table~\ref{table:sdssmulttable} shows that the width of the $\Delta g$ distribution produces comparable uncertainty in $\left<N_{true}(n_{sat})\right>$ as the Poisson sampling.  Therefore, neglecting the integral in Eqn.~\ref{margdeltag} would cause an underestimation of the errors on the HOD parameters.  In our likelihood calculation we use the histogram of $\Delta g(n_{sat})$ values to estimate $P(\Delta g(n_{sat}))$ in Eqn.~\ref{margdeltag}.\\
\begin{deluxetable}{llllll}
\tabletypesize{\scriptsize}
\tablewidth{0pt}
\tablecolumns{6}
\tablecaption{\label{table:deltagvar} Column 2 shows the average offset  $\Delta g(n_{sat}) = N_{CiC}(n_{sat})-N_{true}(n_{sat})$ over 600 mock catalogs for HOD parameters $\sigma_{log M} = 0.7$, $M_{min} = 8.05 \times 10^{13} M_{\sun}$, $M_{cut} = 4.66 \times 10^{13} M_{\sun}$, $M_1 = 4.95 \times 10^{14} M_{\sun}$, and $\alpha = 1.07$, which are close to the final maximum likelihood parameters reported in \S~\ref{HODconstraints}.  Column 3, $\sigma^2_g$, is our estimate of the variance of $\Delta g$ in a subsample of the simulation box with volume equal to our SDSS subsample, 0.46 (Gpc/$h)^3$.  Comparison with Table~\ref{table:sdssmulttable}, Column 2 (reproduced here in Column 4) shows that this variance is comparable to the expected variance from Poisson sampling of $\left<N_{CiC}(n_{sat})\right>$.  Column 5 is the fraction of true one-halo groups with $n_{sat}$ satellites that are exact matches to a CiC group.  Column 6 shows the fraction of CiC groups that do not exactly match a true one-halo group.} 
\tablehead{
\colhead{$n_{sat}$} & \colhead{$\Delta g(n_{sat})$} & \colhead{$\sigma^2_g$} & \colhead{$N_{CiC,SDSS}$} & \colhead{$f_{exact}$} & \colhead{$f_{contam}$}}
\startdata
1 & 286 & 1238 & 2301.12 & 0.77 & 0.33\\
2 & -12.0 & 288 & 285.86 & 0.56 & 0.42\\
3 & -11.8 & 79.5 & 54.29 & 0.38 & 0.55\\
4 & -5.4 & 27.2 & 20.20 & 0.25 & 0.66\\
5 & -2.36 & 8.3 & 4.13 & 0.15 & 0.77\\
6 & 0.99 & 3.2 & 1.05 & 0.10 & 0.83\\
7 & -0.36 & 1.4 & 1.02 & 0.08 & 0.88\\
8 & -0.18 & 0.6 & 0.02 & 0.05 & 0.92\\
9 & -0.04 & 0.3 & 0 & 0.05 & 0.92\\
10 & -0.04 & 0.1 & 0 & 0.02 & 0.96\\
\enddata
\end{deluxetable}
Table~\ref{table:deltagvar} also reports a measure of the completeness and contamination with which we find groups of size $n_{sat}$.  Column 5 is the fraction of true one-halo groups with $n_{sat}$ satellites that are exact matches to a CiC group.  This is a rather stringent definition, since it excludes groups with only one missing satellite of many, or with only one galaxy from a different halo.  Column 6 shows the fraction of CiC groups that do not exactly match a true one-halo group.  Our optimization mainly focused on groups with only one satellite, since they are most numerous.  Moreover, the CiC parameters were established while still using FoF halo catalogs; there are non-negligible differences in the two-halo term between the two catalogs (see \S\ref{fofkulksec}).  Therefore, we are still optimistic that we can improve the accuracy of our method at $n_{sat} > 1$, which we will address in a later paper.  The issue is of less concern in the current work, since we have calibrated the observed statistics on our mock catalogs.  Large biases are unlikely as long as the mock catalogs are reproducing the salient features of the small-scale clustering and distribution of satellites within dark matter halos.\\

\subsection{HOD Constraints}
\label{HODconstraints}
\begin{figure}
\includegraphics*[scale=0.75,angle=0]{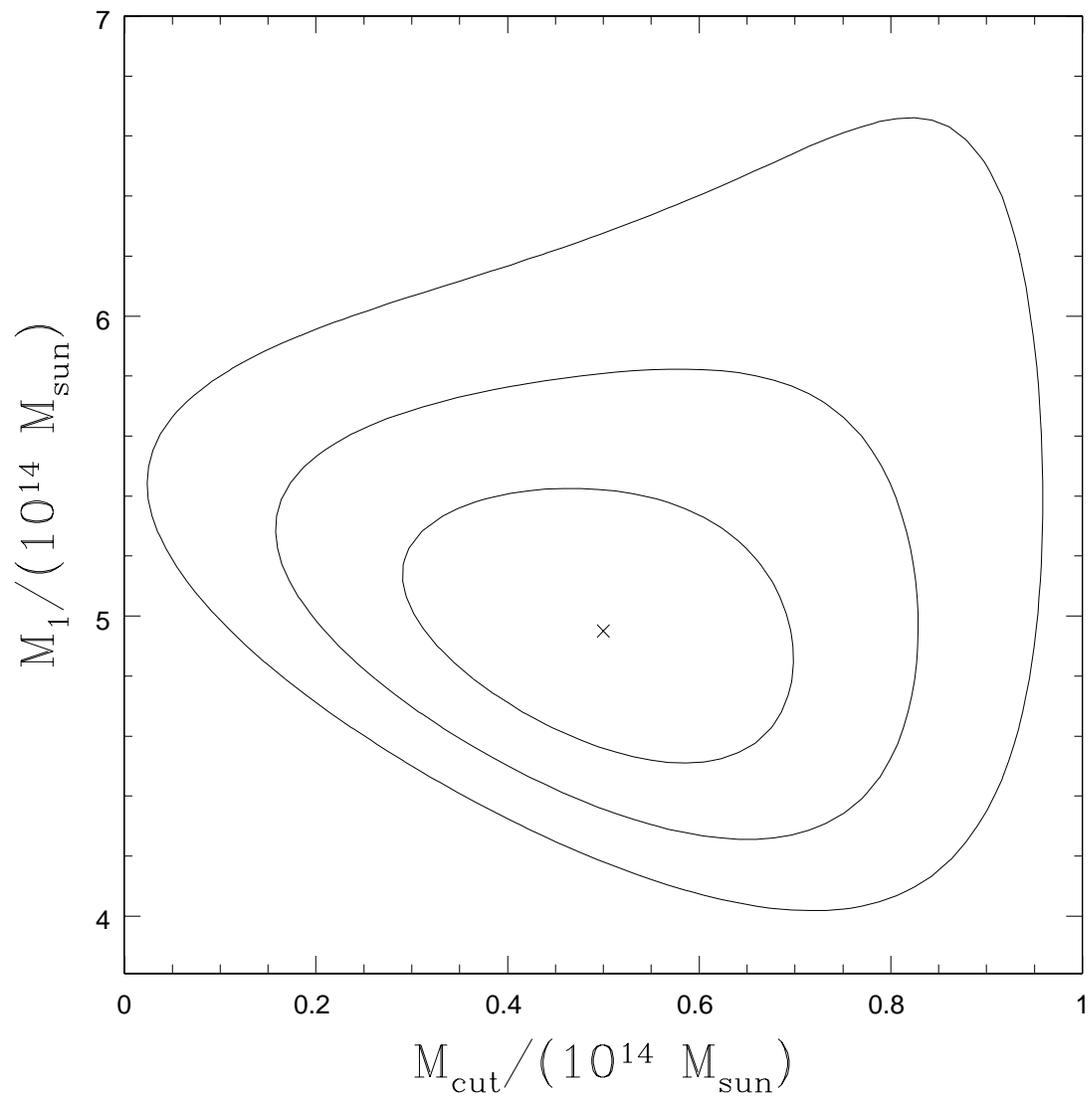}
\caption{\label{fig:sdssM1Mcut}  Contours for $\Delta \ln L = \{-1.15, -3.09, -5.9\}$ in the $M_{1}$ vs. $M_{cut}$ plane after marginalizing over $\alpha$.  For a $\chi^2$ distribution, these contours would enclose 1, 2, and 3$\sigma$ confidence regions.  The cross indicates the maximum likelihood parameter values.}
\end{figure}

\begin{figure}
\includegraphics*[scale=0.75,angle=0]{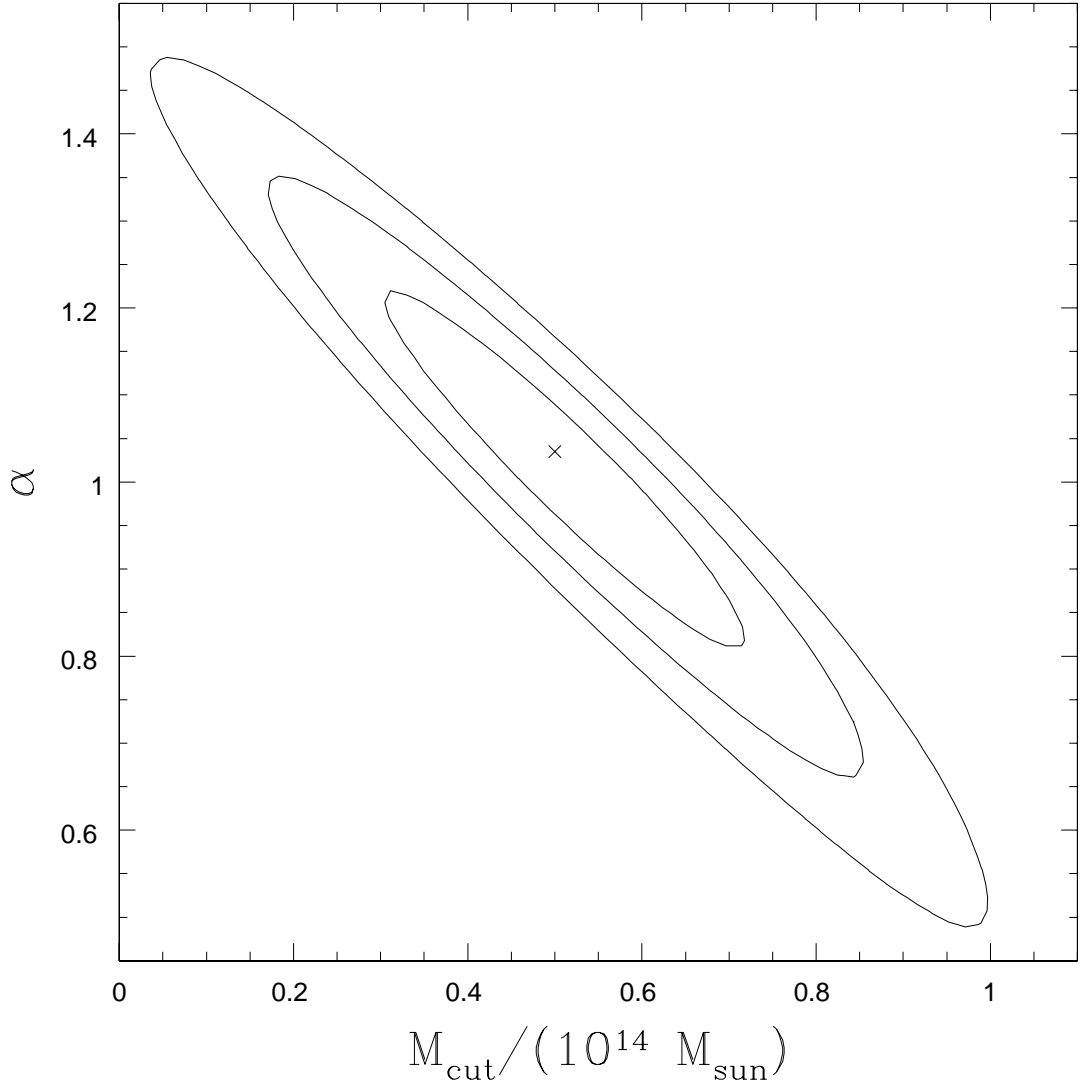}
\caption{\label{fig:sdssalphaMcut} Same as Figure~\ref{fig:sdssM1Mcut}, but for $\alpha$ vs. $M_{cut}$.  $M_{cut}/(10^{14} M_{\sun}) + \alpha$ is tightly constrained, while $M_{cut}/(10^{14} M_{\sun}) - \alpha$ is only weakly constrained.} 
\end{figure}

\begin{figure}
\includegraphics*[scale=0.75,angle=0]{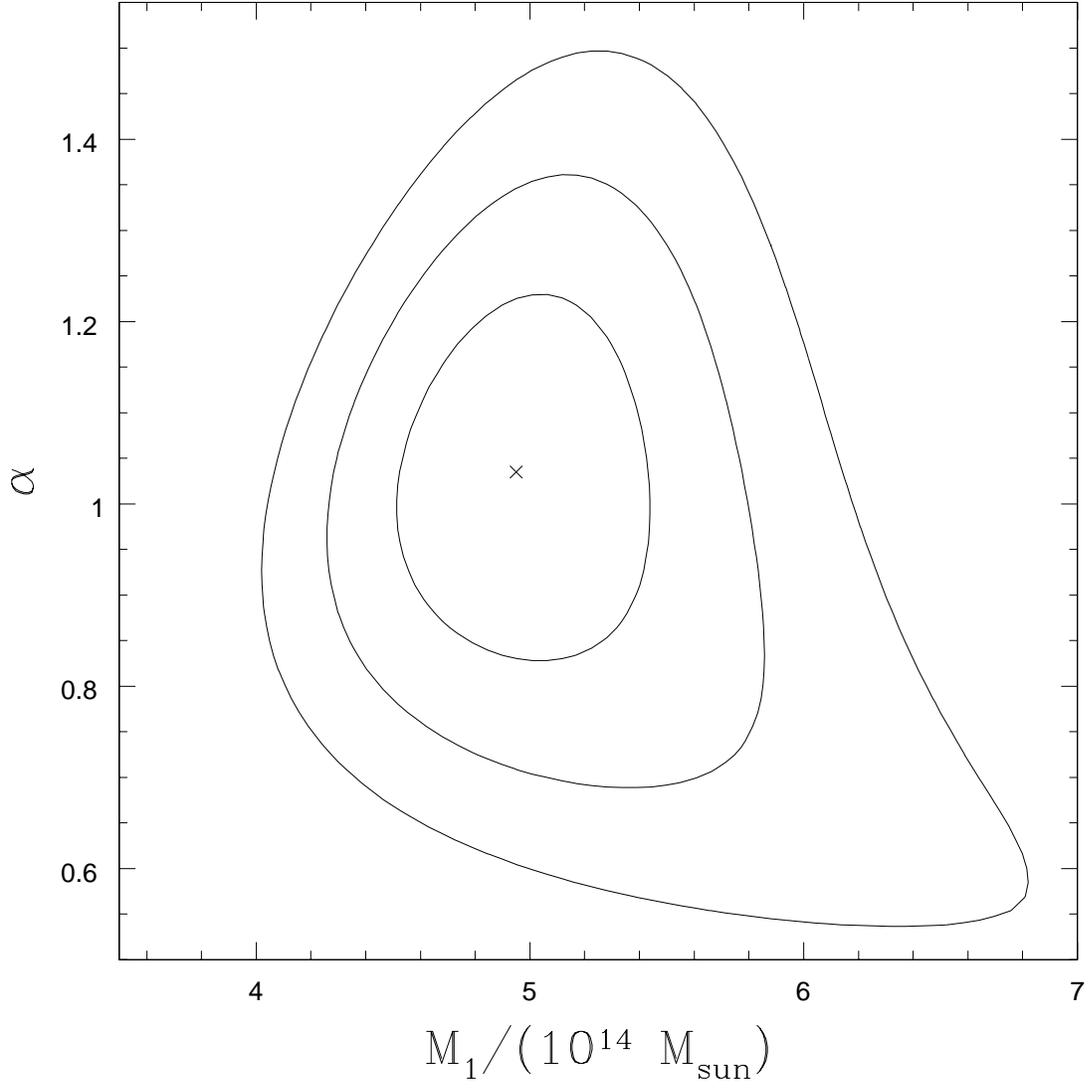}
\caption{\label{fig:sdssalphaM1} Same as Figure~\ref{fig:sdssM1Mcut}, but for $\alpha$ vs. $M_{1}$.} 
\end{figure}

With the width of the $N_{cen}(M)$ distribution fixed at $\sigma_{log M} = 0.7$, we compute the likelihood at each point in the three dimensional space of $N_{sat}$ parameters $\alpha$, $M_{cut}$, and $M_1$ according to Eqns.~\ref{poissondist} - ~\ref{MLeqn}.  We evaluate $M_{min}$ at each HOD point so that $\bar{n}_{LRG}$ remains fixed.  The maximum likelihood HOD parameters and marginalized one-dimensional 68\% and 95\% confidence intervals are $M_{cut} = 5.0^{+1.5}_{-1.3} (^{+2.9}_{-2.6}) \times 10^{13} M_{\sun}$, $M_1 = 4.95^{+0.37}_{-0.26} (^{+0.79}_{-0.53}) \times 10^{14} M_{\sun}$, and $\alpha = 1.035^{+0.10}_{-0.17} (^{+0.24}_{-0.31})$, with $M_{min} = 8.05 \times 10^{13} M_{\sun}$ at the maximum likelihood point.  In Figures~\ref{fig:sdssM1Mcut}, ~\ref{fig:sdssalphaMcut}, and ~\ref{fig:sdssalphaM1} we show two dimensional likelihood contours for $\Delta L = \{-1.15, -3.09, -5.9\}$ after marginalizing over the remaining parameter.  For a $\chi^2$ distribution, these contour values correspond to 1,2, and 3$\sigma$ confidence regions.  We find a strong degeneracy in the $M_{cut}-\alpha$ plane.  We tightly constrain the fraction of galaxies that are satellites, $f_{sat} = 0.0636^{+0.0019}_{-0.0020} (^{+0.0038}_{-0.0039})$ and the parameter combination $\alpha + M_{cut}/(10^{14} M_{\sun}) = 1.53^{+0.039}_{-0.047} (^{+0.080}_{-0.090})$.  In the orthogonal direction, we find only a weak constraint: $M_{cut}/(10^{14} M_{\sun}) - \alpha = -0.54^{+0.32}_{-0.22} (^{+0.60}_{-0.49})$.  In Column 3 of Table~\ref{table:simsmulttable} we list the number of groups with $n_{sat}$ satellites averaged over 20 mock catalogs evaluated at our maximum likelihood HOD, which show good agreement with the SDSS CiC multplicity function; recall that the broad distribution in $N_{CiC} - N_{sat}$ introduces noise in addition to the Poisson term.\\

\begin{figure}
\includegraphics*[scale=0.75,angle=0]{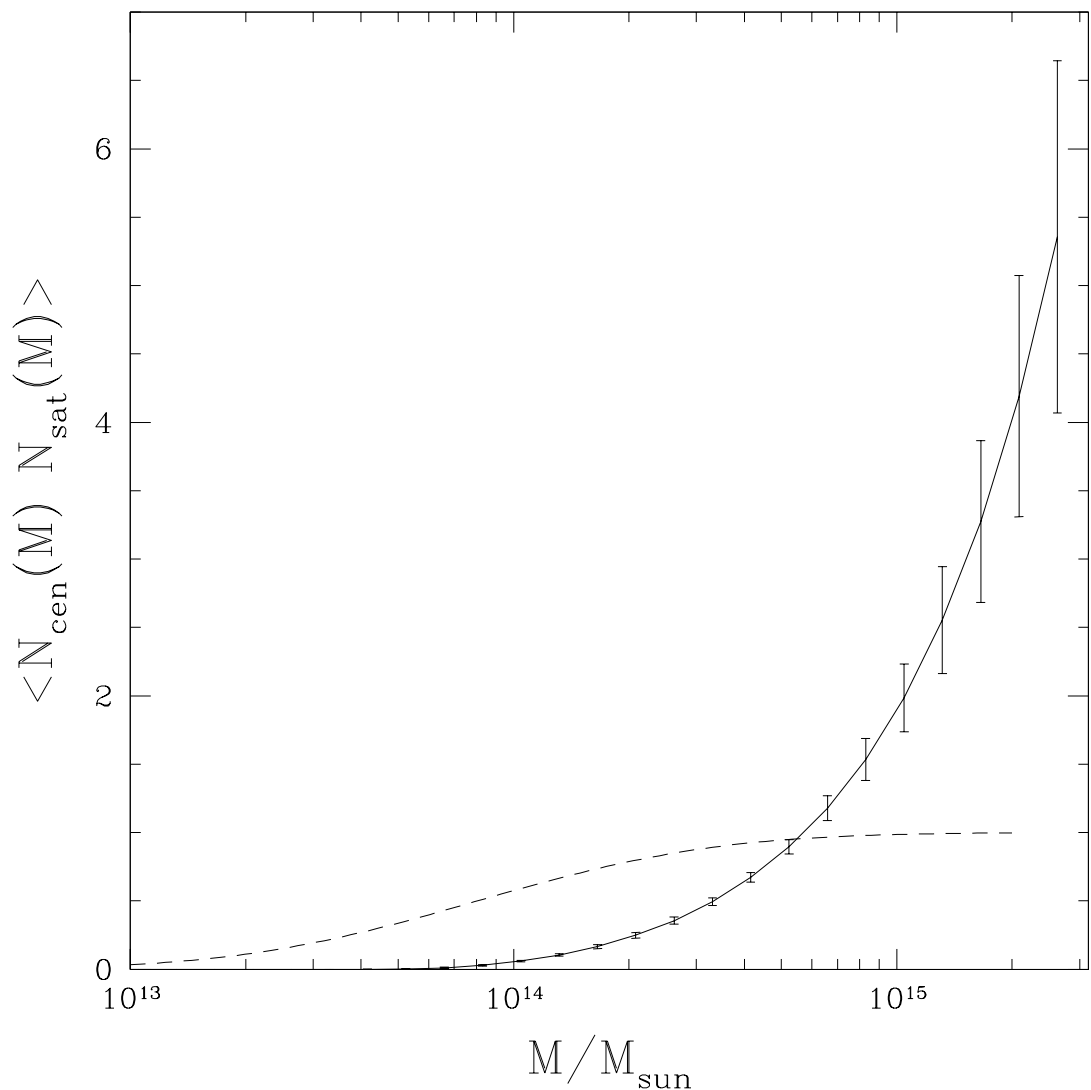}
\caption{\label{fig:avgHOD} The dashed curve ranging from 0 to 1 shows the $N_{cen}(M)$ term for the maximum likelihood HOD; it should vary only slightly with HOD since the satellite fraction is well constrained by our model, and we hold $\sigma_{log M}$ fixed.  The solid curve shows $\left<N_{sat}(M) N_{cen}(M)\right>$, with the error bars computed by Eqn.~\ref{varHODeqn}.}
\end{figure}

\begin{figure}
\includegraphics*[scale=0.75,angle=0]{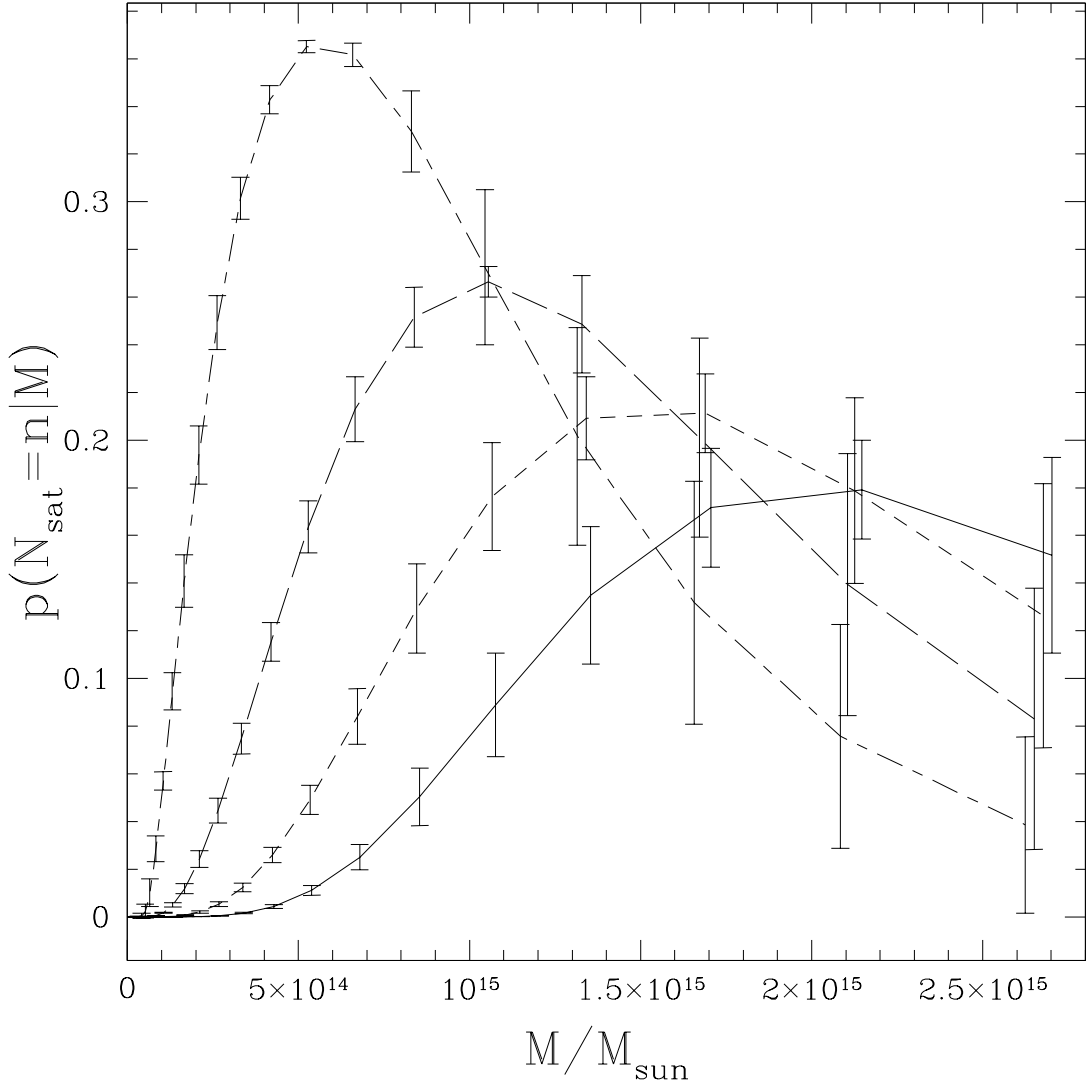}
\caption{\label{fig:pNn} The probability $P(n_{sat} = 1 | M)$ (long-short dashed curve), $P(n_{sat} = 2 | M)$ (long dashed curve), $P(n_{sat} = 3 | M)$ (short dashed curve), $P(n_{sat} = 4 | M)$ (solid curve) vs. halo mass $M/M_{sun}$.  Error bars are computed by Eqn.~\ref{varHODeqn}.  Though $P(n_{sat}|M)$ are evaluated at identical values of $M$, points are slightly staggered for clarity.} 
\end{figure}

\begin{figure}
\includegraphics*[scale=0.75,angle=0]{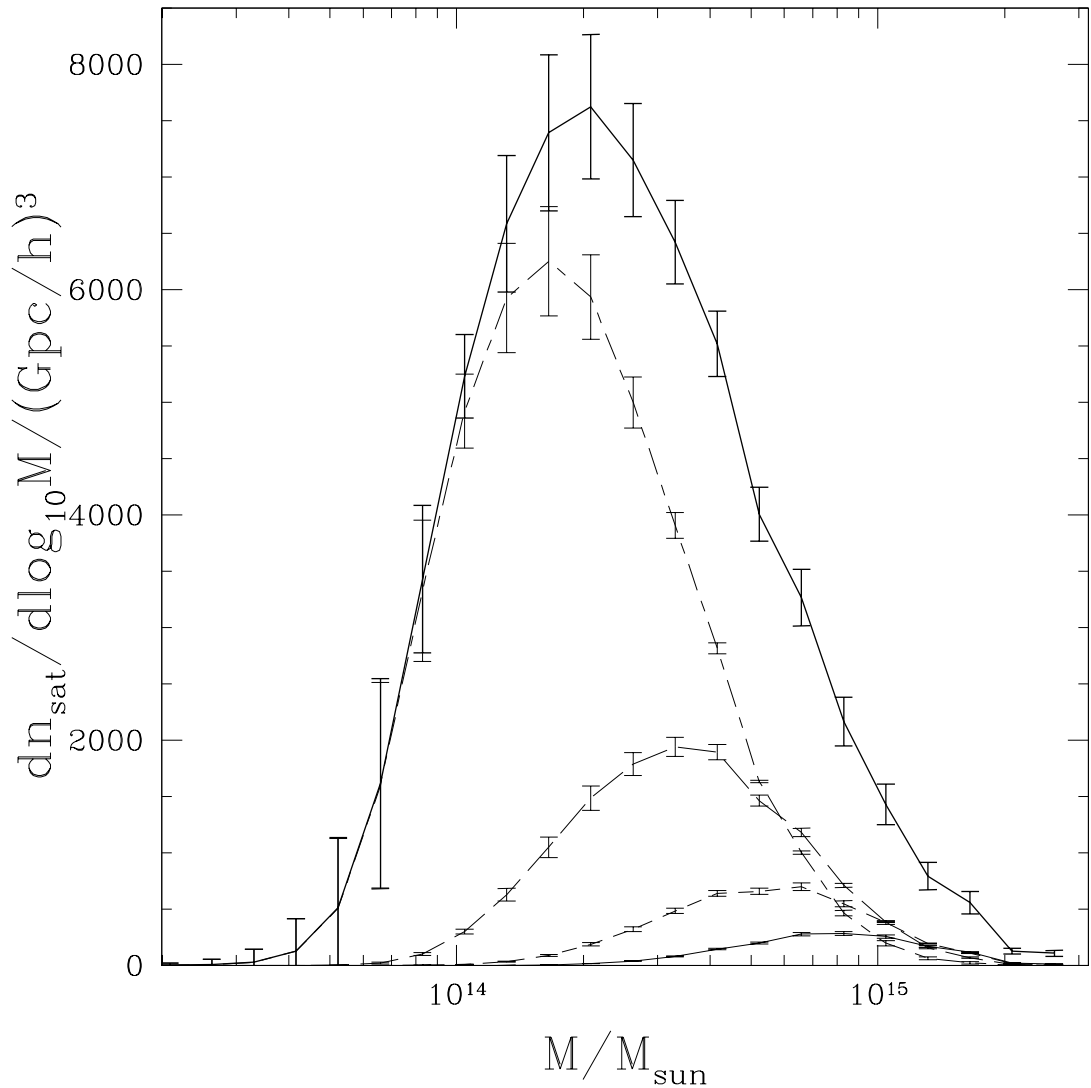}
\caption{\label{fig:dNndlogM} The total number density of satellite galaxies occupying a halo of mass $M$ per $log_{10} M$ (thick solid upper curve).  The long-short dashed curve shows the number density of satellites occupying halos with $n_{sat} = 1$, long dashed curve for satellites in halos with $n_{sat} = 2$, short dashed curve for $n_{sat} = 3$, and solid curve for $n_{sat} = 4$.  Error bars are computed by Eqn.~\ref{varHODeqn}.} 
\end{figure}

In Figures~\ref{fig:avgHOD} through ~\ref{fig:dNndlogM}, we show the implications of these results for the distribution of both satellite galaxies as a whole and groups containing $n_{sat}$ satellites as a function of halo mass.  Figure~\ref{fig:avgHOD} shows the mean and rms (Eqns.~\ref{avgHODeqn} and ~\ref{varHODeqn}) of the expected number of satellites, $\left<N_{cen}(M)N_{sat}(M)\right>$ for several halo masses.  $N_{sat}(M)$ is less well constrained at the high mass end, where the number density of halos is low.  Furthermore, we have not included cosmic variance of halo counts in this analysis; we have only used the mass function from our simulation.  By comparison with Figure~\ref{fig:dNndlogM}, we see that $N_{cen}(M) < 1$ even when $M$ is large enough to host satellite galaxies.  Thus the tight constraints on the low mass end of $N_{sat}(M)$ are dependent on the accuracy of our parametrization of $N_{cen}(M)$.  Figure~\ref{fig:pNn} shows the probability that a halo of mass $M$ hosts $n_{sat}$ satellites.  Again, the constraints are weaker as $M$ increases and $n(M)$ decreases.  Figure~\ref{fig:dNndlogM} shows the distribution of all satellite galaxies as a function of halo mass, as well as the distribution of satellites in groups with $n_{sat} = 1,2,3$ and 4.  The width of these distributions is comparable to the difference in mean halo mass as function of $n_{sat}$; this is qualitatively in line with the large scatter seen in $N_{LRG}$ vs. $M_{200}$ in \citet{ho/etal:2007}.  Attempts to measure the mass of dark matter halos using LRG groups as a tracer should expect a broad distribution.\\
\clearpage
\subsection{Projected Correlation Function $w_p(r_p)$}
\label{projcorr}
\citet{zehavi/etal:2005a} and \citet{masjedi/etal:2006} measure the projected correlation function $w_p(r_p)$ for this sample on small and intermediate scales:
\begin{equation}
w_p(r_p) = 2 \int_0^{\pi_{max}} d\pi \xi(r_p,\pi).
\end{equation}
We follow \citet{zehavi/etal:2005a} and set $\pi_{max} = 80$ Mpc/$h$, which is large enough to include most correlated pairs and produce stable estimates of $w_p(r_p)$.  \citet{masjedi/etal:2006} recover missing fiber collision pairs by computing $w_p(r_p)$ by cross correlation between the SDSS spectroscopic and imaging samples.  They also correct for photometric biases of close galaxy pairs, which can introduce incompleteness of pairs with separation $r_p \lesssim 0.1$ Mpc/$h$.\\

\begin{figure}
\includegraphics*[scale=0.75,angle=0]{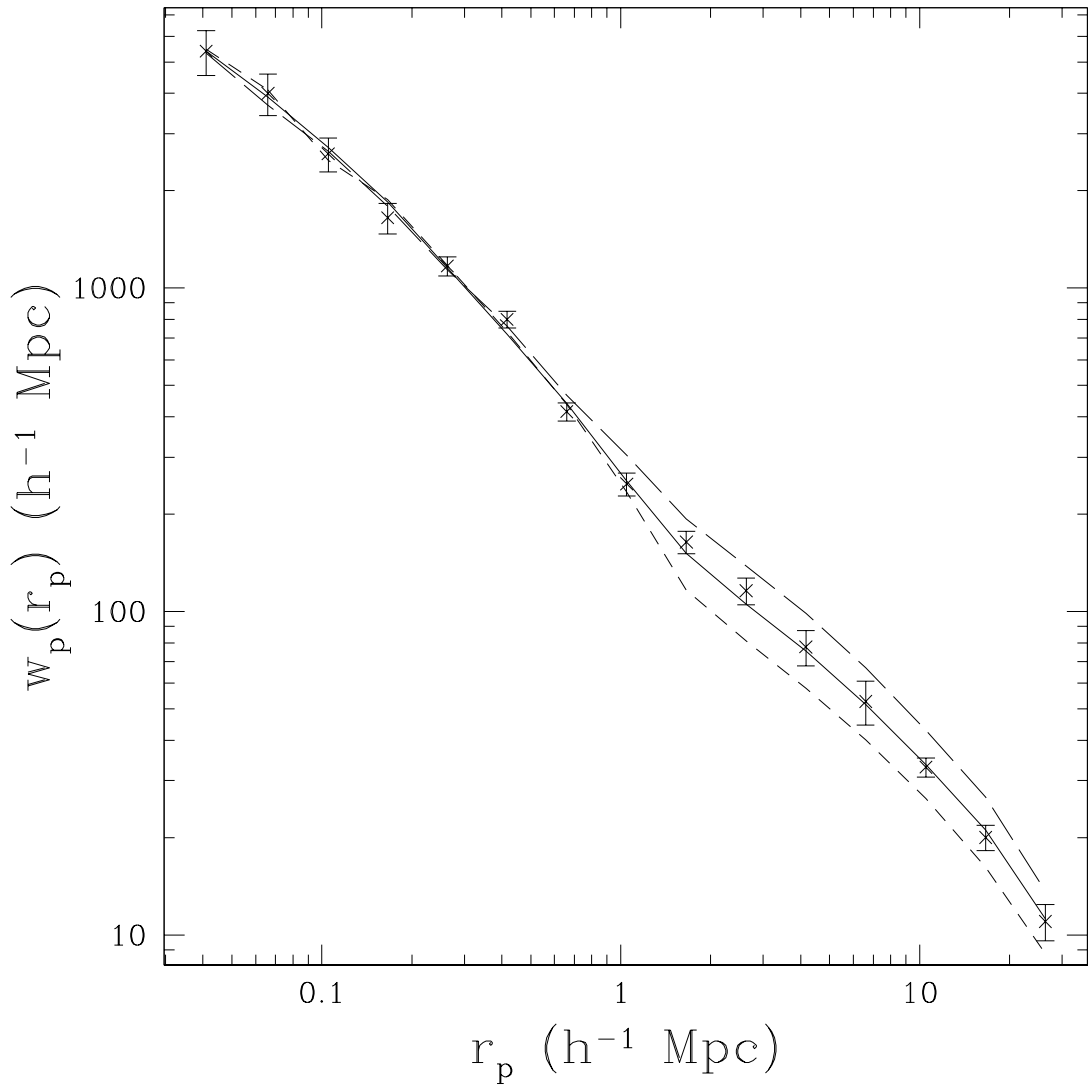}
\caption{\label{fig:wprp1} The projected correlation function  $w_p(r_p)$ vs. projected separation $r_p$, both in Mpc/$h$.  The points with error bars taken from \citet{masjedi/etal:2006}.  The solid curve is an average over 20 mock catalogs at the maximum likelihood HOD parameters presented in the text for $\sigma_{log M} = 0.7$.  The \citet{masjedi/etal:2006} data was used in our analysis only to optimize the value of $\sigma_{log M}$, which sets the large scale clustering.  The striking agreement shows that using SO halo catalogs we can simultaneously reproduce the CiC group multiplicity function and the clustering on scales $\sim 1-20$ Mpc/$h$.  For comparison, we produce a catalog with $\sigma_{log M} = 0.2$ (long dashed curve) and $\sigma_{log M} = 1.3$ (short dashed curve).  Both catalogs match the observed clustering at $r_p \lesssim 0.8$ Mpc/$h$ where we have CiC constraints, but disagree with the observations on larger scales.}
\end{figure}

We present the projected correlation function $w_p(r_p)$ averaged over 20 mock catalogs produced with our SO halo catalog using our maximum likelihood HOD in Figure~\ref{fig:wprp1}.  We find excellent agreement with the measurements of \citet{masjedi/etal:2006}.  Using the diagonal error bars reported in \citet{masjedi/etal:2006}, we find $\chi^2 = 7.5$ for the outer 15 points.  There is substantial discrepancy with the inner 3 points at $r_p = 0.01, 0.016, 0.026$ (not shown in Fig.~\ref{fig:wprp1}); $\chi^2 = 29$ for all 18 points.   The discrepancy is not surprising since these small distances are comparable to the force resolution of our simulation.  Though the CiC method relies primarily on pairs with $r_p \leq 0.8$ Mpc/$h$, our mock catalogs reproduce the features of the observed $w_p(r_p)$ by adjusting a single parameter $\sigma_{log M}$ to match the large scale ($\sim 20$ Mpc/$h$) bias probed by $w_p(r_p)$.  Note that a sharp transition from 0 to 1 for $N_{cen}(M)$ can be ruled out with confidence.  Figure~\ref{fig:wprp1} shows $w_p(r_p)$ for catalogs with $\sigma_{log M} = 0.2$, 0.7, and 1.3 for comparison.  All three catalogs match the observed clustering at $r_p \lesssim 0.8$ Mpc/$h$ where we have CiC constraints, but only catalogs with $\sigma_{log M} \sim 0.7$ match the observed clustering on $\sim 2-20$ Mpc/$h$ scales, the regime where two-halo pairs dominate.\\ 
\clearpage
\subsection{Comparison with Results using FoF Catalogs and Previous Works}
\label{fofkulksec}
\begin{figure}
\includegraphics*[scale=0.75,angle=0]{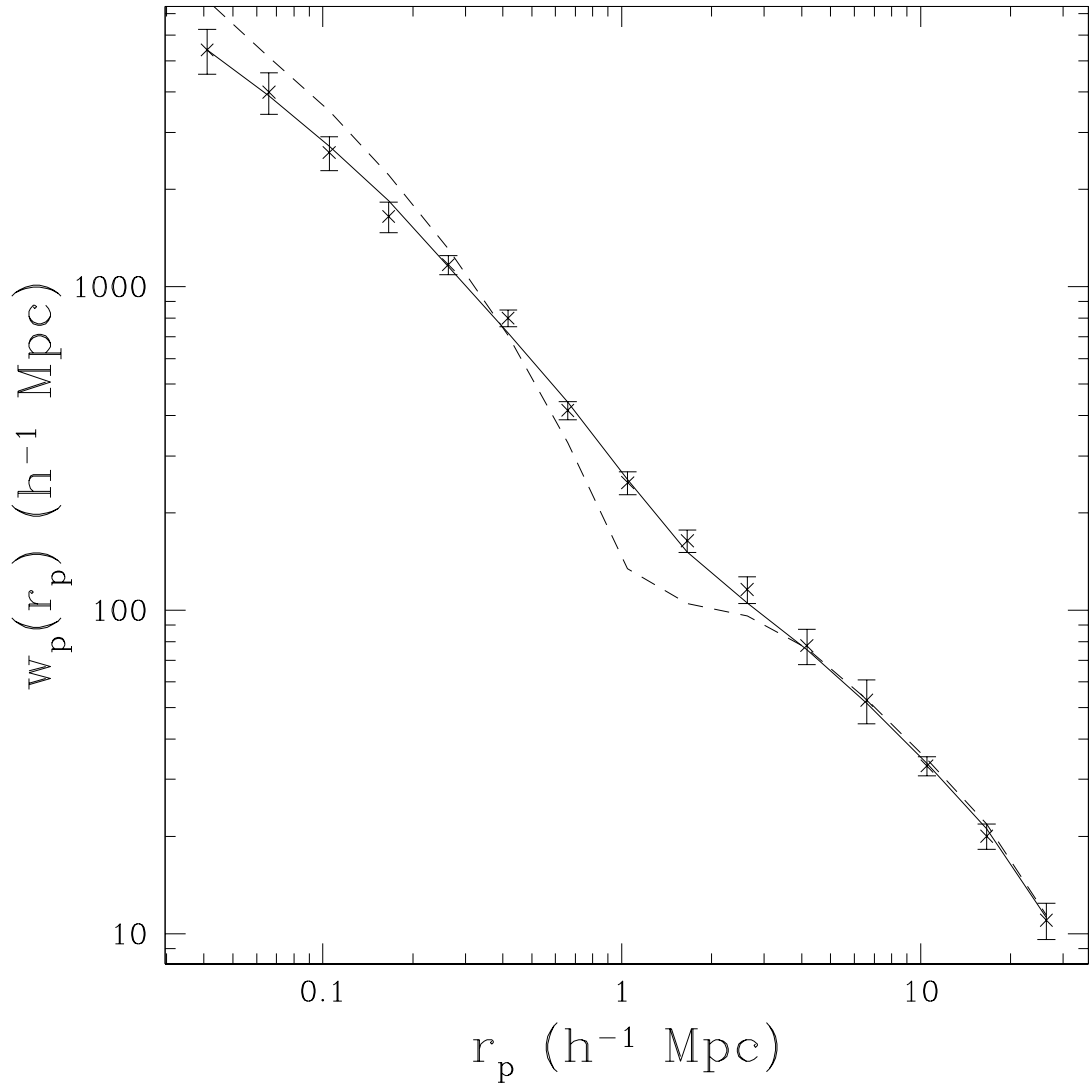}
\caption{\label{fig:fofcompare} The projected correlation function  $w_p(r_p)$ vs. projected separation $r_p$, both in Mpc/$h$.  The points with error bars from taken from \citet{masjedi/etal:2006}.  The solid curve is an average over 20 mock catalogs at the maximum likelihood HOD parameters presented in the text for $\sigma_{log M} = 0.7$ and using the SO halo catalog.  The short dashed curve shows $w_p(r_p)$ averaged over 5 mock catalogs generated by populating an FoF halo catalog.  The FoF halo catalog has a severe suppression of two-halo pairs at separations of $\sim 1$ Mpc/$h$, equal to the half of the total number of pairs in the one-halo regime, $r_p \lesssim 0.8$ Mpc/$h$.}
\end{figure}

We also generate mock LRG catalogs based on a FoF halo catalog produced with linking length $b=0.2$.  For this catalog we use virial masses as defined in \citet{bryan/norman:1998}.  Setting $\sigma_{log M}=0.5$, we are able to match both the SDSS multiplicity function (Column 4 of Table~\ref{table:simsmulttable}) and the large scale projected correlation.  However, Figure~\ref{fig:fofcompare} shows that the FoF catalog does not reproduce the observed $w_p(r_p)$ at $r_p \sim 1$ Mpc/$h$.  Using the diagonal error bars only presented in \citet{masjedi/etal:2006}, this catalog has $\chi^2 = 104$ for the last 15 points (the mock catalog based on SO halos has $\chi^2 = 7.5$).  The large discrepancy between the FoF and SO-based mock catalogs arises from the FoF algorithm's tendency to link nearby halos.  The SO algorithm explicated in \citet{tinker/etal:2008} allows halo overlap in order to correctly separate distinct density peaks likely to host LRGs.  This problem should also be circumvented by algorithms which identify subhalos.\\

If one sums the difference between the observed and FoF catalog $w_p(r_p)$ for the three points near 1 Mpc/$h$, the number of missing pairs in that regime is equal to half the number in the one-halo pair regime, $r_p \lesssim 0.8$ Mpc/$h$.  To match only $w_p(r_p)$ using an FoF catalog, one would therefore need to enhance the number of pairs at separation $\sim 1$ Mpc/$h$ without greatly altering the number of pairs at smaller radii.  In the halo model, this can be accomplished by putting many satellite galaxies in a large halos whose virial radii extend to $\sim 1$ Mpc/$h$, so that satellite-satellite pairs also contribute substantially to $w_p(r_p)$.  In a given halo, satellite-satellite pairs will have larger separations than central-satellite pairs.  This drives $\alpha$ to large values so that the one-halo term can accommodate the missing two-halo pairs in FoF catalogs.  Current LRG HOD results in the literature may therefore be systematically biased if the two-halo term is derived from FoF halo catalogs directly \citep{kulkarni/etal:2007,white/etal:2007,padmanabhan/etal:2008}, based on analytic fits to the HOD calibrated on FoF catalogs from $N$-body simulations \citep{blake/collister/lahav:2007, brown/etal:2008, zheng/etal:2008}, or neglects the scale-dependent bias \citep{wake/etal:2008}.\\

Both \citet{zheng/etal:2008} and \citet{kulkarni/etal:2007} report HOD constraints for the SDSS LRG subsample studied in this paper.  Our tight constraint on the satellite fraction, $6.4 \pm 0.4\%$, is in excellent agreement with the $5.2 - 6.2\%$ found by \citet{zheng/etal:2008}.  However, they find a steeper value of $\alpha$, $1.8$ at $\sigma_8 = 0.8$, than our finding of $\alpha = 1.0^{+0.2}_{-0.3}$ (95\% confidence).  Equivalently, they find $\left <N_{cen}(M) N_{sat}(M) \right> \sim 4$ at $M = 10^{15} M_{\sun}$ while we find $\sim 2$.  This difference is probably due to the difference in our assumed 2-halo correlation function at $\sim Mpc/h$ scales.  \citet{kulkarni/etal:2007} provide their halo mass function, so we are able to make a direct comparison between their predicted multiplicity function and our $N_{CiC}$ measurement.  Column 5 of Table~\ref{table:simsmulttable} shows the expected true multiplicity function from the best fit HOD presented in \citet{kulkarni/etal:2007}, scaled to an SDSS volume.  Their HOD predicts many more groups with $n_{sat} = 1, 2, $ and 3 than are observed; at larger $n_{sat}$ our current CiC method may not recover such large groups.  This HOD also predicts 15 objects in an SDSS volume with $n_{sat} > 15$.  Interestingly, if we fix $M_{cut} = 0$, we also find $\alpha \sim 1.4$ (see Figure~\ref{fig:sdssalphaMcut}).  The cause of the discrepancy here seems to be mainly in $M_1$, the overall normalization of their $N_{sat}(M)$.  While we find $f_{sat} = 0.064$, they report $f_{sat} = 0.17$.  Perhaps their large FoF linking length $b = 0.6$ Mpc/$h$ severely reduces the two-halo clustering in small scales, forcing the one-halo term to compensate.\\

The results of \citet{ho/etal:2007} support our findings for a relatively low cutoff for the number of LRGs per halo.  Figure 12 of \citet{ho/etal:2007} shows a single cluster with 16 LRGs; and all other clusters in their sample have $\leq 11$ LRGs.  We expect this to provide a strict upper bound on our sample, since their photometric sample has a number density $\sim 4$ times larger than the spectroscopic sample we study here.\\
\begin{deluxetable}{lllll}
\tabletypesize{\scriptsize}
\tablewidth{0pt}
\tablecolumns{5}
\tablecaption{\label{table:simsmulttable} Column 1 is the number of satellites in the group.  Column 2 is a copy of Column 3 of Table~\ref{table:sdssmulttable}, our best estimate of the SDSS CiC multiplicity function.  Column 3 shows the CiC multiplicity function averaged over 20 mock catalogs produced with our maximum likelihood HOD using the SO halo catalogs and scaled by the volume ratio $V_{SDSS}/V_{sim}$.  Column 4 shows the CiC multiplicity function for mocks made with the FoF catalog.  Column 5 shows our estimate of the expected $N_{true}$ in $V_{SDSS}$ for the \citet{kulkarni/etal:2007} HOD.}
\tablehead{
\colhead{$n_{sat}$} & \colhead{$N_{\rm SDSS CiC,final}$} & \colhead{$N_{\rm SO ML HOD CiC}$} & \colhead{$N_{\rm FOF}$} & \colhead{$N_{\rm Kulkarni}$}}
\startdata
0 & 40407.56 & 40546.1 & 40542.0 & 37855\\
1 & 2301.12 & 2190.4 & 2265.1 & 5202\\
2 & 285.86 & 323.2 & 301.3 & 998\\
3 & 54.29 & 65.8 & 53.7 & 343\\
4 & 20.20 & 15.2 & 13.4 & 155\\
5 & 4.13 & 4.7 & 4.0 & 82\\
6 & 1.05 & 1.5 & 1.6 & 48\\
7 & 1.02 & 0.2 & 0.92 & 30\\
8 & 0.02 & 0.16 & 0.28 & 21\\
\enddata
\end{deluxetable}
\section{Assessment of the CiC method}
\label{assessCiC}
The HOD parameter constraints reported in \S~\ref{HODconstraints} are for $\sigma_{log M}$ held fixed.  Since the CiC constraints always involve the product $\left<N_{cen}(M) N_{sat}(M)\right>$ and Figure~\ref{fig:avgHOD} shows that $N_{cen}(M) < 1$ in regions where the satellite contribution is non-negligible, the CiC maximum likelihood HOD parameters will vary with $\sigma_{log M}$ to hold $\left<N_{cen}(M) N_{sat}(M)\right>$ relatively fixed.  Moreover, varying $\sigma_{log M}$ changes the average bias of the LRG-occupied halos, which may in turn affect the distribution of $\Delta g(n_{sat})$ in Eqn.~\ref{deltag}.  We expect this to be a smaller effect.  A complete HOD analysis would include $N_{cen}(M)$ parameters when evaluating the likelihood.  We find that using $w_p(r_p)$ on $\sim 20$ Mpc/$h$ to set $\sigma_{log M}$ is sufficient for the purposes of generating mock catalogs.  We hope to explore variations in the functional form of $N_{cen}(M)$ in later work.  We have not explored the dependence of our method on the assumed distribution of satellite galaxies.  \citet{seo/eisenstein/zehavi:2007} find that passively evolving galaxies are more concentrated than the dark matter at low redshift; this should make the CiC group finder more accurate than we found for our mock catalogs.  For the parameters chosen here, the CiC method does not extract exact matches to one-halo groups with high fidelity.  We have circumvented this issue by careful calibration on our simulated catalogs, and we expect to refine the method in the future to improve the accuracy of finding groups with more than two LRGs.\\

Because the CiC method can simultaneously reproduce the large scale clustering of LRGs and match the higher order statistics in the density field probed by the CiC group multiplicity function, we expect it to be a useful tool both for HOD constraints and the production of good mock catalogs.  Since the CiC method only uses information on scales where the one-halo term dominates, our method is less sensitive to knowing the two-halo correlation function on $\sim$ Mpc/$h$ scales.  We have demonstrated that the deficit of two-halo pairs at $\sim 1$ Mpc/$h$ is problematic for HOD fitting methods using 2 and 3-pt statistics and FoF catalogs because the HOD must accommodate the missing two halo pairs at this separation, and thus predict groups containing many more LRGs than are observed.\\

Finally, the low number density of the LRGs makes them particularly well-suited for this method.\\

\section{Conclusions}
\label{conc}
The low number density of SDSS LRGs allows us to partially separate LRG pairs occupying the same dark matter halo from pairs occupying distinct dark matter halos.  Candidate one-halo pairs are identified using simple cuts in the transverse and LOS separations.  We group these pairs using the FoF algorithm to compute the CiC group multiplicity function.  We measure the CiC group multiplicity function for the subsample of SDSS LRGs satisfying $-23.2 < M_{g} < -21.2$ and $ 0.16 < z < 0.36$, carefully accounting for the effects of fiber collisions and survey boundaries, holes, and incompleteness.\\

In order to derive HOD constraints from our measurement, we calibrated the relation between the CiC and true one-halo group multiplicity functions using mock LRG catalogs.  The variance about the mean relation is comparable to the Poisson sampling variance of the CiC multiplicity function and must be properly accounted for in the maximum likelihood parameter estimation.\\

The CiC group multiplicity function places strong constraints on the satellite LRG HOD, $N_{sat}(M)$.  When we fix $\sigma_{log M} = 0.7$ and $\bar{n}_{LRG} = 10^{-4}$ (Mpc/$h)^{-3}$, the maximum likelihood HOD parameters and marginalized one-dimensional 68\% and 95\% confidence intervals are $M_{cut} = 5.0^{+1.5}_{-1.3} (^{+2.9}_{-2.6}) \times 10^{13} M_{\sun}$, $M_1 = 4.95^{+0.37}_{-0.26} (^{+0.79}_{-0.53}) \times 10^{14} M_{\sun}$, and $\alpha = 1.035^{+0.10}_{-0.17} (^{+0.24}_{-0.31})$, with $M_{min} = 8.05 \times 10^{13} M_{\sun}$ at the maximum likelihood point.  We tightly constrain the satellite fraction to $f_{sat} = 0.0636^{+0.0019}_{-0.0020} (^{+0.0038}_{-0.0039})$.  The projected correlation function $w_p(r_p)$ of mock catalogs derived from an SO halo catalog is an excellent agreement with the measurements of \citet{masjedi/etal:2006} and \citet{zehavi/etal:2005a} when the large scale clustering is used to fix $\sigma_{log M}$.  Fig.~\ref{fig:fofcompare} shows that FoF halo catalogs have a severe deficit of pairs at $\sim 1$ Mpc/$h$.  In \S~\ref{fofkulksec} we point out that methods using $w_p(r_p)$ with an analytic estimate of the two-halo term calibrated using FoF halos will severely overestimate the number of satellites and the maximum expected one-halo group size.  Our measured CiC group multiplicity function rules out the best fit HOD from \citet{kulkarni/etal:2007}.\\

Despite the increased complexity of our approach and necessary calibration using simulations, we have produced high quality mock catalogs which reproduce both higher order statistics in the density field and the features of the projected correlation function.  These mock catalogs will be used in a forthcoming paper to study the large scale structure statistics of our CiC groups.\\

\section{Acknowledgments}
Daniel Eisenstein provided our final LRG sample along with excellent advice regarding the SDSS analysis and Michael Blanton provided the inverse random catalogs and Tycho2 catalog.  We thank Jeremy Tinker for excellent discussions regarding FoF vs. SO halo finding algorithms as well as kindly shared his SO halo-finding code, which was used to produce the halo catalogs in this paper.  Rachel Mandelbaum and Charlie Conroy provided excellent guidance and suggestions in the initial stages of this project.  Paul Bode kindly provided the simulation results and assistance.\\

Funding for the Sloan Digital Sky Survey (SDSS) has been provided by the Alfred P. Sloan Foundation, the Participating Institutions, the National Aeronautics and Space Administration, the National Science Foundation, the U.S. Department of Energy, the Japanese Monbukagakusho, and the Max Planck Society. The SDSS Web site is http://www.sdss.org/.\\
The SDSS is managed by the Astrophysical Research Consortium (ARC) for the Participating Institutions. The Participating Institutions are The University of Chicago, Fermilab, the Institute for Advanced Study, the Japan Participation Group, The Johns Hopkins University, Los Alamos National Laboratory, the Max-Planck-Institute for Astronomy (MPIA), the Max-Planck-Institute for Astrophysics (MPA), New Mexico State University, University of Pittsburgh, Princeton University, the United States Naval Observatory, and the University of Washington.\\

BAR gratefully acknowledges support from the National Science Foundation Graduate Research Fellowship.  This project was supported by NSF Grant 0707731.  DNS thanks the APC (Universite de Paris VII) for its hospitality while this work was completed.
\bibliographystyle{apj}
\bibliography{apj-jour,/u/breid/bibfile/bethbibfile}\end{document}